# Impacts of thrusting, extensional faulting, and glaciation on cratering records of Pluto's largest moon Charon: Implications for the evolution of Kuiper belt objects

Hanzhang Chen, and An Yin* (ayin54@gmail.com)

*Department of Earth, Planetary, and Space Sciences, University of California, Los Angeles, California 90095-1567, USA*

**Abstract**

A first-order question in the studies of the Solar System is how its outer zone known as the Kuiper belt was created and evolved. Two end-member models involving coagulation vs. streaming instability make different predictions about the cumulative size-frequency distribution (SFD) of Kuiper Belt Objects (KBOs), which are testable by the cratering history of KBOs. Among all of the imaged KBOs, Pluto's largest icy moon Charon appears to preserve the largest size range of seemingly undisturbed craters with their diameters (D) ranging from <1 km to >220 km. Current work shows that Charon's craters with D<10-20 km are less than those expected by the coagulation mechanism, but whether this is an artifact of post-cratering modification of smaller craters is unknown. We address this issue by conducting systematic photogeological mapping and performing detailed landform analysis using the highest resolution images obtained by the New Horizons spacecraft that reveal the presence of a range of differentiable terrains on Charon. The most important findings of our work include (1) truncation and omission of large craters (diameters > 30-40 km) and their crater rim ridges along the eastern edges of several north-trending, eastward-convex, arcuate ranges in Oz Terra of the northern encountered hemisphere, (2) lobate ridges, lobate knob trains, and lobate aprons resembling glacial moraine landforms on Earth, (3) dendritic channel systems containing hanging valleys, and (4) locally striated surfaces defined by parallel ridges, knob trains, and grooves that are >40-50 km in length. The above observations and the topographic dichotomy of Charon's encountered hemisphere can be explained by a landscape-evolution model that involves (i) a giant impact that created the Vulcan Planitia basin and the extensional fault zone along its northern rim, (ii) a transient atmosphere capable of driving $N_2$-ice glacial erosion of the water-ice bedrock and transporting water-ice debris to sedimentary basins, (iii) regional glacial erosion and transport of earlier emplaced impact ejecta deposits from the highlands of Oz Terra into the lowland basin of Vulcan Planitia, (iv) syn-glaciation north-trending thrusting interpreted to have been induced by Charon's despinning, and (v) the development of a water-ice debris cover layer over subsurface N2 ice below Vulcan Planitia during global deglaciation. The infilling of the Vulcan Planitia could have been accompanied by cryovolcanism. The extensive modification of impact craters means that the crater size-frequency distributions from Charon should serve only as a lower bound when used to test the formation mechanism of Kuiper belt objects.

**1. Introduction**

Understanding the formation of the Kuiper belt is central to quantifying the dynamic evolution of the Solar System (Nesvorný, 2018). A starting point of this endeavor is to establish the cumulative size-frequency distribution (SFD) and evolution of Kuiper Belt Objects (KBOs), which can be used to test competing models of planet formation during the initial <100-Myr development of the Solar System (Youdin and Goodman, 2005; Schlichting et al., 2013; Abod et al., 2019; Morbidelli et al., 2021). To this end, the current understanding of the KBO size distribution comes from two independent sources of observations: (i) telescopic surveys (Schlichting et al., 2012; Fraser et al., 2014), and (ii) crater records of recently imaged KBOs by the New Horizons spacecraft (Robbins et al., 2017; Singer et al., 2019; Spencer et al., 2021). Among the imaged KBOs, Pluto's largest icy moon Charon appears to preserve the largest size range of seemingly undisturbed craters with their diameters ($D$) ranging from <1 km to >220 km. Detailed crater mapping and size counting show a significant deficit $D$<10-20 km when compared to the projected trend of the KBO sizes determined by the telescopic surveys (Singer et al., 2019). This deficit is expressed by a gentler slope for craters with $D$ = 2-13 km than the slope for craters with $D$ = 13-30 km in the plot of cumulative crater numbers vs. crater diameters (Robbins et al., 2017; Singer et al., 2019). The deficit of smaller craters implies collisional disequilibrium of the KBOs that have sizes smaller than 1-2 km across. This inference is contrary to the coagulation models that predict KBOs to have entered a state of collisional cascade within the first ~100 million years (Myr) of the Solar System history (Stern, 1995; Farinella and Davis, Schlichting, 2011; Schlichting et al., 2013). The discrepancy, however, could be explained by the neglect of ice strength in the early coagulation models (Kenyon and Bromley, 2020), or streaming instabilities during the formation of KBOs (Youdin and Goodman, 2005; Abod et al., 2019). The last explanation was favored by Singer et al. (2019).

The coagulation vs. streaming-instability hypotheses for the KBO formation can be tested by the cratering record on Charon. For example, the reported crater SFD by Robbins et al. (2017) and Singer et al. (2019) could have been affected by later resurfacing events (Morbidelli et al., 2021). In this case, the coagulation mechanism cannot be rejected. On the other hand, if the reported crater SFD has not been disrupted by younger geological events, a streaming-instability mechanism would be supported (Singer al., 2019).

In this study, we systematically address the critical issue of whether the crater record of Charon has been contaminated by later geological processes. This goal was achieved by performing detailed, systematic, and highest-resolution geomorphological mapping up to this study across the entire encounter hemisphere of the satellite. Our mapping was accompanied by detailed landform analyses, and their interpreted formation mechanisms were guided by well-documented and well-understood analogues from Earth, Mercury, and Mars. These efforts show that most craters on Charon have been modified by previously unrecognized geological processes such as thrusting, extensional faulting, glaciation, crater-filling viscous flows, and possible fluvial processes. This new understanding in turn implies that the crater size-frequency data from Charon should be regarded only as a lower bound when used for testing competing models of Kuiper-belt formation.

## 2. Pluto-Charon System

The Pluto-Charon system was interpreted to have been formed by the collision of two similarly sized progenitors within the first 10 Myr of the Solar System history (McKinnon, 1989; Canup, 2005; Canup et al., 2021; McKinnon et al., 2021). The distribution of their spin/orbital periods and the small eccentricity of the Pluto-Charon system favor an explanation that Pluto and Charon were fluid-like at the onset of their formation, which implies that the two bodies had experienced rapid differentiation at the time of their initial formation (Arakawa et al., 2019). Thermal-modeling results show that the early differentiation should lead to the formation of ancient subsurface oceans within the two bodies, and that of the Charon should have been refrozen at ca. 4.0 billion years (Ga) (Beyer et al., 2017) or 2.5-2.0 billion years ago (Ga) (Desch and Neveu, 2017; Bierson et al., 2018) depending on the nature and evolution of the assumed heat sources. Because Pluto has a $N_2$-dominated atmosphere, the shared origin with its largest moon implies that Charon may also have had a primordial $N_2$ atmosphere. This inference was anticipated from theoretical considerations (Trafton et al., 1988) and the current understanding of Kuiper belt history (e.g., Stern et al., 2015). This scenario is likely, because Charon's initial surface temperature is considered to be at ~100 K, possibly induced by the combination of a closer distance to the Sun (Nesvorný, 2018) and the release of heat from the Charon-forming collision (Canup et al., 2021; McKinnon et al., 2021), gravitational accretion (Canup et al., 2021; McKinnon et al., 2021), radiogenic decay (Canup et al., 2021; McKinnon et al., 2021), internal phase changes

(Malamud et al., 2017; Canup et al., 2021; McKinnon et al., 2021), and tidally induced deformation (Rhoden et al., 2020; Conrad et al., 2021; Stern and Trafton, 2008). If the atmospheric and surface vapor pressures were in equilibrium, a surface temperature ($T$) higher than nitrogen melt temperature of 63.15 K would imply surface vapor pressures of >100 μbar for the dominant KBO volatile species such as $N_2$, CO, and $CH_4$ (Fray et al., 2019). The above pressure and temperature conditions on Charon could have resulted in fluvial processes involving $N_2$ liquid. As a result of atmosphere escape, the subsequent surface temperature on Charon after its initial formation should have decreased below the freezing temperature of $N_2$ liquid, which would have favored $N_2$-ice glaciation on the surface of Charon. Glaciation on Charon is possible when considering the occurrence of past and current glaciation on Pluto (Howard et al., 2017) and the fact that the current $N_2$ vapor pressures on Pluto are $10^{-2}$ to $10^3$ μbar inferred from theoretical modeling (Hansen and Paige, 1996) and actual measurements (Gladstone et al., 2016; Bertrand et al., 2018; Johnson et al., 2021), which are comparable to the pressure magnitudes deduced from the theoretical considerations (Trafton et al., 2008).

Contrary to the expected primordial atmosphere over early Charon as inferred by Trafton et al. (1988), the existing literature interprets its landscape to have evolved mainly through cratering, global extension, and cryovolcanism (Beyer et al., 2017; Moore et al., 2015; Beyer et al., 2019; Robbins et al., 2019; Spencer et al., 2021). Although the above inferred geologic processes explain a large number of observations from Charon, they are unable to account for the following observations.

(1) Larger craters ($D$>50 km) consistently lack ejecta blankets and secondary craters (Moore et al., 2016; Robbins et al., 2017, 2019; Protopapa et al., 2021). This observation requires a geological process capable of removing the impact-generated materials and landforms, which has not been discussed in the current literature.

(2) Trough networks dominate Oz Terra on Charon (Moore et al., 2016; Beyer et al., 2016), the assigned normal-faulting mechanism (Beyer et al., 2016) does not explain converging dendritic-like trough networks and locally highly curvilinear trough traces (see more details below).

(3) The origin of the topographic dichotomy boundary between Oz Terra and Vulcan Planitia (Figs. 1A and 1B) in the encounter hemisphere of Charon has been attributed to a giant impact (Malamud et al., 2017) or zonal extension (Beyer et al., 2019). Both hypotheses have issues with the current observations: the impact hypothesis does not explain the lack of impact breccias whereas the extension hypothesis does not explain why faulting is localized along the rim of Vulcan Planitia.

(4) No explanation has been offered for the exposure of higher volume contents and larger grain sizes of crystalline water ice that are uniquely restricted along north-trending ranges in Oz Terra (Protopapa et al., 2021). As noted below, the exposures of the crystalline water ice are exclusively associated with arcuate ranges that truncate $D>50$ km craters.

The above unresolved issues hint additional geological processes contributing the geomorphological development of Charon. For example, the lack of impact breccias around the largest craters on Charon requires erosion and transport processes, while the inconsistency in explaining the dichotomy boundary implies more complicated geologic processes in shaping the morphologies of Charon.

**3. Data and Methods**

Images collected by the Long-Range Reconnaissance Imager (LORRI; Cheng et al., 2008) and Multi-spectral Visible Imaging Camera (MVIC) (Reuter et al., 2008; Howett et al., 2017) on board of the New Horizons spacecraft are used in our geomorphological mapping (Schenk et al., 2018a). Landform analyses are assisted by the use of digital elevation models (DEMs) created by Schenk et al. (2018a). The resolutions of the images used in this study vary from ~154 m/pixel to ~1460 m/pixel. Meanwhile, the DEMs used in this study have a uniform map-view resolution of 300 m/pixel (Schenk et al., 2018a; Robbins et al., 2019).

The shape and distribution of each mapped landform was determined using both the satellite image and DEM at the same location. By doing so, we are able to minimize the artifacts created by the spatial variation in pixel scales, solar incident angles, viewing directions of three-dimensional objects, and emission angles (Schenk et al., 2018a). All DEMs used in this study are geographically rectified and shown in a cylindrical projection. According to Schenk et al. (2018a), the horizontal resolution of the DEMs is ~300 m/pixel. In contrast, the vertical resolution is ca. 100-400 m for most of the area covered in this study except its rim regions where resolutions degraded to ~1500 m (see figure 4 in Schenk et al., 2018a). Note that all the topographic profiles shown in this study have vertical uncertainties of <500 m (Figs. 2 and 9).

Geological mapping was conducted on the highest resolution images available, and the resulting maps were drafted using the Adobe Illustrator$^{TM}$. Analogue examples from the Earth were taken from images available from published work and Google Earth$^{TM}$, whereas analogue examples from Mars and Mercury are taken from NASA's publically accessible data.

Images used for mapping were first processed using Adobe Photoshop $^{TM}$ for the best display of their brightness and contrast. Because of this, caution should be taken by the readers when comparing the images used in this study with their unprocessed counterparts taken directly from NASA's database. Because craters on Charon were mapped in detail by Robbins et al. (2017), Singer et al. (2019), and Robbins and Singer (2021), our mapping does not repeat this effort. Instead, our mapping focuses only on selected craters that illustrate different cratering styles and different relationships between the mapped craters and the surficial landforms and/or surficial materials next to the craters.

Interpreting the formation mechanism of a single element in a geological system often yields highly non-unique solutions, and such a narrowly focused investigation rarely answers the central scientific questions. Because of this, we do not offer interpretations of each terrain after its description as commonly done by researchers in the planetary geology community. Rather, we adopted the land system approach (e.g., Brodzikowski and van Loon, 1987; Yin et al., 2021) aiming at constructing the first-order landscape-evolution model of Charon that counts for the spatiotemporal relationships among all facets and all elements in the analyzed geomorphological system as constrained by our systematic and detailed mapping as presented below.

## 4. Results of Geomorphologic Mapping

The fundamental goal of our work is to constrain the spatiotemporal relationships among all geomorphological features in Charon's encounter hemisphere based on systematic landform analysis. Fig. 1A is an enhanced color view of Charon that combines blue, red, and infrared images taken by the New Horizons spacecraft's Ralph/Multispectral Visual Imaging Camera (MVIC) (Moore et al., 2016). Fig. 1B is a digital elevation model of Charon's encounter hemisphere at a horizontal resolution of 300 m/pixel overlying the Charon New Horizons LORRI MVIC Global Mosaic 300 m v1 (Schenk et al., 2018a). Fig. 1C is a geomorphological map of Charon's encounter hemisphere created in this study. Description of terrain units is listed on the column to the right.

Although our geomorphological mapping is built upon the early studies of Charon (Moore et al., 2016; Beyer et al., 2017, 2019; Robbins et al., 2019; Beddingfield et al., 2019), our work differs in the following two ways. First, we separate elongated linear and curvilinear depressions (i.e., troughs) into two types: one bounded by flat plains (e.g., feature 6 in Fig. 1B) and the other bounded by rift-shoulder-like ridges (e.g., feature 3 in Fig. 1B) (cf., Beyer et al., 2017, 2019). Second, we split the previously mapped two landform units by Beyer et al. (2017, 2019) and Robbins et al. (2019) in Oz Terra into ten differentiable terrains (Fig. 1C). This detailed terrain classification allows us to make more comprehensive and self-consistent interpretations of the landform-formation mechanisms. Description of the mapped units, their correlations with earlier mapped terrains, and their temporal relationships (i.e., chronostratigraphy) are summarized in Figs. 1C and 1D.

*4.1 Map Units*

We describe the landforms units in our regional geomorphological map (Fig. 1C) based on their formation ages from the oldest to the youngest.

**Unit $ct_1$.** This unit is concentrated in Oz Terra and represents the oldest and largest-sized craters characterized by degraded crater-basin morphologies such as lack of central uplifts and rim ridges. They are also marked by a total absence of ejecta-apron deposits as noted by Moore et al. (2016).

**Unit $ct_2$.** This unit occurs both in Oz Terra and Vulcan Planitia and represents the intermediate-aged and intermediate-sized group of craters that are characterized by layered ejecta-apron deposits commonly displaying rampart-like morphologies as those on Mars (cf., Carr et al., 1977; Weiss and Head, 2013).

**Unit *ct₃*.** This unit represents the youngest-aged and smallest-sized group of craters that occur both in Oz Terra and Vulcan Planitia. They are characterized by light-toned radial ejecta rays similar to those on the lunar surface (cf., Morse et al., 2018).

**Unit *rv*.** This unit represents ridge-bounding valley zones that are bounded by linear scarps. The valleys may have a graben-like morphology with their rims bounded by rift-should-like ridges.

**Unit *rt*.** This is a ridged terrain characterized by linear and parallel ridges. This terrain also exposes polygonal depressions and the terrain surface is locally scattered by angular knobs.

**Unit *lr*.** This terrain is characterized by east-trending linear ranges that have scarp-defined range fronts and bound graben-like elongated basins mapped as unit *rv* as described above. This unit occurs only along the southern margin of Oz Terra directly north of Vulcan Planitia.

**Unit *cp*.** This unit represents a highland catered-plain terrain that consists of the oldest, degraded craters (unit *ct₁*) and polygonal networks of troughs in Oz Terra and a highland region bounding the western end of Vulcan Planitia.

**Unit *tr₁*.** This terrain is characterized by polygonal networks of troughs that dominate the landscape of Oz Terra. Note that some troughs in this terrain are truncated by north-trending range-bounding scarps.

**Unit *th*.** This is a highland terrain that consists of an older set of wide (20-30 km), long (>200-300 km), and deep (3-6 km) troughs. The troughs are either terminated at or trend parallel to the north-trending arcuate ranges that also occur in this terrain. The arcuate ranges are cut across by a younger set of smaller troughs, which display converging dendritic patterns and terminate locally at trench-like linear depressions along the northern rim of Vulcan Planitia.

**Unit *ar*.** This unit represents a geomorphological element characterized by north-trending, arcuate range fronts in the troughed highland terrain (unit *th*).

**Unit $ct_F$.** This unit represents craters whose central uplifts are buried by younger ridged materials. The type location is a crater on Oz Terra that shows degraded rim ridges. The same crater truncates a north-trending arcuate range. The crater rim ridge is locally cut across by troughs that extend from the crater-bounding plains to the crater interior.

**Unit $tr_2$.** This unit represents a younger set of dendritic troughs. The troughs are characterized by their termination at closed basins and/or trench-like depressions along the northern edge of Vulcan Planitia. Note that this unit was mapped as a radially trending graben system by Beyer et al. (2019) (see their fig. 16).

**Unit *cb*.** This unit denotes closed basins or trench-like linear/curvilinear depressions along the northern edge of Vulcan Planitia. The rims of the basins and trenches are mostly flat as parts of the bounding plains.

**Unit *sp*.** This terrain is characterized by smooth plains that are cut by networks of linear and curvilinear grooves in Vulcan Planitia.

**Unit *am*.** This unit consists of polygonal-shaped mounds that are ~30-40 km across. The mounds are rimmed by similarly shaped moats that have asymmetric profiles: steepening towards the mounds and shallowing towards the bounding plains. This unit only occurs in Vulcan Planitia and was mapped as "the mons terrain" by Robbins et al. (2019).

**Unit *fb*.** This unit marks a distinctive geomorphological element in Vulcan Planitia, which is expressed as funnel-like, polygonal-shaped, angular-cornered depressions. The depressions have characteristic downward-steepening profiles, whereas the rims of the depressions display cylindrical shapes in the horizontal directions. The angular corners of the depressions exhibit antiformal ridges and synformal grooves. The former were referred to as "broad warps" by Beyer et al. (2019) and Robbins et al. (2019). The funnel-like depressions correlate roughly with "depressed material 3" unit of Robbins et al. (2019).

**Unit *rp*.** This is a ridged-plain terrain best developed along the northern margin of Vulcan Planitia. The unit is intermingled with southward-convex, lobate-shaped ridges that are dotted by angular knobs along the ridge crests.

**Unit *sf*.** This is a smooth-surfaced terrain in Vulcan Planitia, which hosts ridges, grooves, and knobs.

**Unit *la$_S$*.** This unit represents a geomorphological element characterized by its smooth-surfaced, lobate-apron morphology. This type of landforms occur only in the southeastern corner of Oz Terra.

**Unit *ln*.** This unit consists of parallel linear grooves and ridges exposed on larger planar ridge flanks that bound the northern edge of Vulcan Planitia. The grooves in this terrain are interrupted by smaller knobs within the linear depressions. Some of the linear depressions in this terrain are defined by linked pits. This unit was mapped as "grooves" by Beyer et al. (2019) and "scarp crests" by Robbins et al. (2019).

**Unit *rf*.** This unit denotes a basin that has a rough-surfaced floor. The rough texture of the basin floor is locally defined by north-trending, evenly spaced ridges. The basin is bounded by a linear scarp in the south and Dorothy crater in the northwest.

**Unit *kb*.** This unit is characterized by the occurrence of angular to circular knobs, knob trains, and parallel curvilinear ridges.

**Unit *la$_R$*.** This unit denotes a rough-surfaced, lobate-apron plateau terrain bounded by the largest troughs in Oz Terra. The plateau surface locally displays parallel ridges and furrows trending north and parallel to those exposed in the rough-surfaced terrain (*rf*).

**Unit *lb*.** This unit occurs along the northern margin of Vulcan Planitia that slopes gently southward. The unit is characterized by lobate ridges locally dotted by angular and circular knobs along ridge crests. The lobate ridges are convex southward in the direction of the regional slope.

**Unit *k*.** This unit denotes the mappable knobs that have sizes from 1 to 10 km. The knobs exhibit rectangular, polygonal, circular, and oval shapes. The knobs are scattered across Vulcan Planitia but locally present on Oz Terra. The mapped knobs correlate locally with "the mons unit" of Robbins et al. (2019) in the Vulcan Planitia area.

**Unit *pt$_O$* and Unit *pt$_V$*.** These two units are characterized by pitted surfaces in Oz Terra denoted by the subscript "O" and Vulcan Planitia denoted by the subscript "V". This unit correlates approximately to the mottled terrain of Robbins et al. (2019).

**Unit *gv*.** This unit denotes networks of rill-like grooves exposed in Vulcan Planitia. The grooves show *en echelon* alignments and Y-shaped triple junctions. The groove landform mapped in this study is the same as that described in Moore et al. (2016), Beyer et al. (2019), and Robbins et al. (2019).

**Unit *ls*.** This unit denotes lobate-apron-shaped landforms that occur at the base of linear ridge scarps. These features were interpreted as landslides by Beddingfield et al. (2019).

*4.2 Morphologies and Cross-Cutting Relationships*

The most dominant terrain in Oz Terra is the cratered-plain unit (*cp* in Fig. 1C), which is the oldest terrain that hosts degraded craters (unit *ct$_1$*) characterized by the lack of clearly defined impact ejecta deposits, central uplifts, and rim ridges. The cratered-plain terrain also hosts polygonal trough networks dominated by north-trending troughs (unit *tr$_1$*). These troughs are ~10s km wide, >100s km long, and ~3-5 km in local reliefs against the bounding ranges. The troughs display undulating longitudinal profiles (Fig. 2A), U-shaped cross sections, and flat floors (Fig. 1B). Note that Fig. 2B is severely exaggerated in the vertical direction, with the V/H ratio = 12. Because of this distortion, the cross section appears to be V-shaped.

The cratered-plain terrain also hosts north-trending, arcuate-shaped ranges that are ~200 km long, 40-60 km wide, and up to ~2 km in the maximum relief (labelled as *'ar'* in Fig. 1C). The ranges consistently display steeper eastern flanks, gentler western flanks (Figs. 2E), and a westward-concave map-view shape (Figs. 1B and 1C).

The polygonal troughs and north-trending arcuate ranges are mapped together as the troughed highland terrain (unit *th* in Fig. 1C), locally overlain by a knobbed terrain (unit *kb* in Fig. 1C) that consists of linear and curvilinear ridges, linear and curvilinear knob trains, and patches of scattered pits (labelled as '*pt_O*' in Fig. 1C). Fig. 3A shows key morphological features and their spatiotemporal relationships in the troughed highland terrain (*th*). In this image, east-facing scarps (feature 1a) bounding the north-trending arcuate ranges have longer, gentler west-sloping flanks (feature 2) and shorter, steeper eastern range flanks. Note that the ranges are separated by parallel northwest-trending troughs (feature 3a in Fig. 3A). An arcuate range front (feature 1b in Fig. 3A) truncates both crater rims (feature 4 in Fig. 3A) and an older set of east-trending troughs (feature 5 in Fig. 3A). The north-trending ranges themselves are crosscut by a younger dendritic network of troughs (feature 6 in Fig. 3A). The ranges are also truncated (see feature 1b in Fig. 3A) by a younger crater (feature 7 in Fig. 3A) filled by a ridged-plain material. A degraded crater with a dark-material apron (feature 8 in Fig. 3A) is crosscut by an east-trending trough (feature 3b in Fig. 3A).

Fig. 3B is a zoom-in view of an east-facing scarp zone shown in Fig. 3A, which consists of a single scarp trace in the south (feature 1a) and multiple scarp traces (features 1b and 1c) in the north. The scarp zone truncates craters (e.g., features 2 in Fig. 3B) and troughs (feature 3 in Fig. 3B), but the zone itself is crosscut by younger craters (e.g., features 4 in Fig. 3B) and superposed over by a younger dendritic trough system (feature 5). The scarp-bounded range crest displays a set of range-front-parallel ridges (feature 6 in Fig. 3B). The scarp-bounded west-sloping range flank is draped over by a younger crater surrounded by proximal darker-toned and distal lighter-toned ejecta deposits (feature 7 in Fig. 3B).

The arcuate ranges (feature *ar* in Fig. 1C) are locally cut across by east-trending and westward-converging dendritic networks of troughs (unit *tr₂*) (Fig. 3C), which are shorter and narrower than the orthogonal troughs mapped as unit *tr₁*. Note that in Fig. 3C, the dendritic network of troughs (unit *tr₂*) exhibits hanging-valley morphologies (feature 1) and terrace-rise-like scarps (features 2a-2c) along the margin of a meandered, steep-walled trough (feature 3). However, like polygonal troughs, the trunk channel of the dendritic troughs *tr₂* displays undulating longitudinal profiles (Fig. 2C) and U-shaped cross sections when the vertical exaggeration shown in Fig. 2D is removed.

The north-trending arcuate ranges (unit *ar* in Figs. 1C and 1D), locally crosscut by younger craters (*ct*_F in Figs. 1C and 1D), are terminated by the east-trending ridge-bounding valley zone (unit *rv* in Figs. 1C and 1D). The truncated arcuate ranges do not cut across the valley zone, but an arcuate range south of the valley zone is present along the western end of Vulcan Planitia (feature 1 in Fig. 1B). The right-lateral offset of two north-trending ranges north (feature 2 in Fig. 1B) and south (feature 1 in Fig. 1B) of the valley zone is ~400 km.

Valleys in the eastern segment of the valley zone (unit *rv*) are bounded by rift-shoulder-like rim ridges (feature 3 in Fig. 1B). The ridges change trend from northeast (feature 4 in Fig. 1B) to northwest (feature 5 in Fig. 1B) around the northeastern corner of Vulcan Planitia. In contrast, valleys in the western segment of the valley zone (unit *rv*) are bounded by plains without rim ridges (feature 6 in Fig. 1B; cf., feature 3 in Fig. 1B). Note that the topographic profiles of the valley-bounding, east-trending ridges differ from those of the north-trending, accurate ranges in that the former have higher reliefs and steeper range fronts (Figs. 2E and 2F).

The oldest smooth-plain terrain (unit *sp* in Fig. 1C) is dotted by isolated knobs (feature '*k*' in Fig. 1C), polygonal-shaped mounds (unit *am* in Fig. 1C), funnel-like depressions (unit *fb* in Fig. 1C), variously shaped pits (unit *pt$_V$* in Fig. 1C), and craters surrounded by layered ejecta deposits (unit *ct$_2$* in Fig. 1C). The largest funnel-like depression (~70 km in the longest dimension) with a triangular map-view shape (feature 7 in Fig. 1B) occurs in the topographic transition zone between the eastern edge of Vulcan Planitia and western margin of the linear ridge terrain (unit *rt* in Fig. 1C). The smooth plains terrain is crosscut by northwest-trending ridges (labelled as '*cr*' Fig. 1C) hosted by the ridged plains terrain (unit *rp* in Fig. 1C). The ridges are in turn truncated by the smooth-surfaced lobate-apron terrain (unit *la$_S$* in Fig. 1C), which is crosscut by the youngest rough-surfaced terrain (unit *la$_R$* in Fig. 1C). Unit *la$_R$* itself defines a lobate-shaped plateau (Fig. 3E; also see feature 8 in Fig. 1B). The plateau terrain differs from arcuate ranges (i.e., feature '*ar*' in Fig. 1C) in that its surface lacks troughs.

Fig. 3D displays key landforms across the topographic boundary zone between Oz Terra and Vulcan Planitia. Dorothy crater in the northeastern part of the image (feature *i*) is characterized by a smooth-surfaced basin floor and lies next to the rough-surfaced lobate-apron terrain (*la$_R$*) on Oz Terra (feature *ii*). The smooth-surface lobate-apron terrain (feature *iii*) extends from Oz Terra to Vulcan Planitia (feature *iv*). Features 1a-1c in Fig. 3D mark the contact between Dorothy crater and unit *la$_R$*, whereas features 2a and 2b in Fig. 3D marks the contact between unit *la$_R$* and the smooth-surfaced lobate-apron terrain (unit *la$_S$* and feature *iii*). Features 3a-3b in Fig. 3D are examples of mapped knobs in unit *la*R, which are up to >20 km in the longest dimension, whereas features 3c and 3d in Fig. 3D are examples of knobs mapped in unit *la*s with smoother surfaces than those in unit *la*R (cf. feature 3b in Fig. 3D). Note that a knob is draped over a south-facing scarp (feature 3e in Fig. 3D). Feature 4a in Fig. 3D marks a zone of northwest-trending linear troughs mapped as unit *tr*$_2$ that are crosscut by a northeast-trending linear depression (feature 5a in Fig. 3D) and a similarly trending scarp (feature 5b in Fig. 3D). The troughs (unit *tr*$_2$) and their bounding walls are draped over by unit *la*s (features 4b-4d in Fig. 3D). Unit *la*s hosts south-convex lobate aprons (features 6a, 6b and 6c in Fig. 3D), south-convex lobate ridges (features 6d and 6e), north-trending linear ridges and grooves (features 7a and 7b in Fig. 3D), and craters filled by flat smooth-surfaced materials (feature 8 in Fig. 3D; also see description of this same feature in Robbins et al., 2019). A north-trending ridge (feature 7c on the left-central edge of the image in Fig. 3D) exposed on the surface of a south-facing scarp terminates downward at a lobate apron (feature 8 in Fig. 3D). The lack of a corresponding breakaway topographic feature rules out the apron landform to be a landslide. Feature 9 in Fig. 3D shows a possible degraded crater cut by a south-facing scarp, which likely represents the trace of a normal fault. Feature 10 in Fig. 3D represents the contact between the smooth-surfaced lobate-apron unit (features *iii* and *iv* in Fig. 3D) and the ridged plain terrain (feature *v* in Fig. 3D).

    Feature 11 in Fig. 3D are flow-like landforms that cut across crater rims and extend into the crater-basin floors. Layered impact deposits for one crater are deformed by northwest-trending ridges (feature 11a in Fig. 3D), whereas layered impact deposits for another crater (feature 11b in Fig. 3D) are truncated (feature 12b in Fig. 3D) and overprinted (feature 12c in Fig. 3D) by a zone of parallel ridges (*cr*) in the ridged plains terrain (*rp*). The boundary between ridged-plain terrain (*rp* and feature v) and the smooth-surfaced lobate-apron terrain (unit *la*s and feature *vi*) is

transitional (see feature 13 in Fig. 3D). This contact is crosscut by younger grooves (feature 14 in Fig. 3D). In the southwestern corner of the image, northwest-trending smooth-surfaced linear ridges (feature 16a in Fig. 3D) are locally superposed by knobs (feature 3f in Fig. 3D). In contrast, ridges with the same trend directly to the southeast display rougher surfaces (feature 16b in Fig. 3D). A partially filled crater lies at the base of a range bounding scarp (feature 17 in Fig. 3D).

The geomorphological relationship described above are summarized in a geomorphologic map shown in Fig. 3E. Within the four younger units that are superposed over the eastern valley zone (*vz* in Fig. 3E), we note that (1) all lobate-apron landforms are convex southward, (2) craters with layered ejecta deposits have their rims locally breached and floors partially filled by viscous-flow materials (labelled as '*v*' in Fig. 3E), and (3) layered ejecta deposits are locally superposed by ridges (labelled as '*cr*' in Fig. 1C) that are crosscut by younger polygonal networks of grooves (labelled as '*gv*' in Fig. 1C). Note that all other units shown in Fig. 3E are defined in Fig. 1C.

*4.3 Degraded Craters*

Craters mapped as unit *ct*$_1$ and unit *ct*$_2$ display variably degraded morphologies (Figs. 4 and 5). The post-cratering modification is expressed by partial breaching of crater rims, partial infilling of crater basins, and superposition of grooved, ridged, hummocky-textured, and smooth-surfaced materials as detailed below.

Fig. 4A shows a crater with a partially preserved rim ridge marked as feature 1 along its northern and southern edges. In contrast, the western and eastern rims of the same crater are bounded by flat plains where the rim ridge is missing (feature 2). The crater basin is cut across by a series of northwest-trending ridges and grooves (feature 3), and the crater floor and crater-bounding plains share the same hummocky surface texture. The latter relationship requires a post-cratering resurfacing process that created the shared surface morphology. A smaller crater to the southeast of the crater mentioned above is partially superposed by a linear ridge that cuts across the rim of the smaller crater (feature 4 in Fig. 4A; also see details in Fig. 4B).

Fig. 4B shows a crater with its western rim ridge breached and cut across by an east-trending knob train (feature 1). Parallel east-trending ridges are also present on the crater floor (feature 2). A central mound is present on the crater floor (feature 3).

Fig. 4C shows a crater with its western and eastern rims superposed by east-trending ridges and grooves (features 1 and 2). A wider ridge cutting across the crater rim has a dumbbell shape characterized by the occurrence of higher mounds at the two ends (feature 3).

Fig. 4D shows a zoom-in view of the smaller crater shown in Fig. 4A. The crater rim is breached in three places (features 1, 2, and 3), which are expressed by the superposition of 2-5 km wide linear grooves over the crater edge. The crater floor displays curvilinear ridges that terminate at the ends of the crater-rim-cutting grooves (feature 4).

Fig. 4E shows a crater that is breached by northeast-trending grooves and ridges across its western and eastern rims (features 1 and 2). The crater floor displays a central uplift with a well-defined peak (feature 3).

Fig. 4F shows a crater with its western rim superposed by an east-trending, 8-12-km wide, >30 km long ridge (feature 1). The flat top of the ridge is superposed by narrower ridges that are parallel to the general trend of the bigger, hosting ridge (feature 2). The wider ridge splits into two branches as it enters the crater floor (features 3a and 3b).

Fig. 4G shows a sheet-like landform that is superposed over the northwestern rim of a crater (feature 1). The sheet-like landform partially enters the crater basin (feature 2). The sheet-like landform with an abrupt termination margin appears to have been sourced from a smaller, rampart-style impact crater from the west (feature 3).

Fig. 4H shows a crater that has a central peak and ring-shaped ridges and grooves on the crater floor. The eastern rim (feature 1) and western rim (feature 2) of the crater are draped over by younger ridges. The eastern ridge originates from a mound and narrows as it extends into the crater floor.

Fig. 4I shows another crater that has a central uplift and well-preserved northern and southern rim ridges. However, its eastern rim (feature 1) and western rim (feature 2) are breached by ~7-km wide, round-topped ridges that extend into the interior of the crater floor.

Fig. 4J shows an east-trending ridge system that cuts across a quasi-triangular-shaped crater, whereas Fig. 4K shows a circular depression (feature 1) with a smooth and flat floor that may represent a partially filled crater basin. Note that in Fig. 4K, feature 2 denotes a circular-rim ridge that bounds a discontinuous circular groove. The latter appears to be the relics of a filled crater.

Figs. 4L and 4M show a DEM image of Schenk et al. (2018a) along with a corresponding optical image from the same area. In both images, features 1, 2, and 3 mark linear depressions that cut across the crater rim. In the same images, feature 4 denotes a tongue-shaped landform superposed over the crater wall and partially the crater floor.

Figs. 5A-5B show a crater with its southern margin (feature 1) breached and crosscut by a series of northeast-trending ridges (feature 2) along a gap between the rim ridges. Figs. 5C-5D show a crater (feature 1) with its southwestern margin buried by a younger sheet-like landform (feature 2). The same images also show a partially preserved semi-circular ridge (feature 3) that bounds a flat-floored circular basin (feature 4). Note the rim ridge of the basin is breached in the northeast (feature 5). The circular shaped basin (feature 4) could have originated from an impact, and its flat floor and the breached rim ridge require the crater to have been degraded by post-impact processes.

Figs. 5C-5D show a semi-circular ridge (feature 6) bounding a quasi-circular depression (feature 7) with a complex floor morphology due to the presence of superposed impact craters (feature c). Note that the circular basin marked as feature 7 does not have the northeastern rim, which is occupied by a plain connecting with the basin floor (feature 8). Note that a southwest-pointing lobate ridge (feature 9) cuts across the northeastern rim of a smaller crater. Distinctive landforms are also displayed in Figs. 5C and 5D. These include arête-like ridges (feature A), dendritic network of troughs (feature D), hanging-valley-like landforms (H), flat-topped lobate aprons (feature la), pyramid-like peaks (feature P), and U-shaped valleys (feature U).

The superposed post-cratering landforms in Figs. 4 and 5 are typically 5-10 km in dimension. This observation implies that craters older than the superposed landforms and smaller than 5-10 km in size could have been completely or partially buried. On the other hand, craters larger than 5-10 km in size would have been better preserved on the surface of Charon.

*4.4 Chronological Sequence of Landform Units*

The sequence of geologic events is defined by the cross-cutting relationships discussed above and can be summarized as the following contact relationships as summarized in Fig. 1D. The contact relationships are numbered from the oldest to the youngest in age for the clarity of description below.

**Contact 1.** The Vulcan Planitia basin (not its surficial deposits) is interpreted to be the oldest geomorphological feature, because its margin represented by the ridge-bounding valley zone (unit *rv*) and linear-ridge terrain (unit *lr*) are crosscut and/or superposed by all other mapped landform units.

**Contact 2.** The troughed highland terrain (unit *th*) featuring north-trending arcuate ranges (unit *ar*) is superposed on top of the cratered-plain terrain (unit *cp*), older troughs (unit $tr_1$), and degraded craters (unit $ct_1$) that lack ejecta aprons and display partially preserved crater rims.

**Contact 3.** The north-trending arcuate ranges (unit *ar*) are crosscut by a crater (unit $ct_F$) with the locally breached rim ridge, younger dendritic troughs (unit $tr_1$), and younger elongated trench-like depressions (unit *cb*) bounded by the smooth plains terrain (unit *sp*) hosting polygonal-shaped mounds (unit *am*), funnel-like basins (unit *fb*), and knobs of various shapes and sizes (unit *k*).

**Contact 4.** Parallel linear and curvilinear ridges (unit *cr*) in the ridged plains terrain (unit *rp*) that cut across the smooth plains terrain (unit *sp*).

**Contact 5.** Smooth-surfaced terrains with (unit $la_S$) or without (unit *sf*) lobate landforms are superposed over the ridged plains terrain (unit *rp*).

**Contact 6.** A rough-floored basin terrain cuts across the linear-ridge terrain (unit *lr*) and the smooth-surfaced lobate-apron terrain (unit $la_S$).

**Contact 7.** The rough-surfaced lobate-apron terrain (unit $la_R$) truncates landforms in unit $la_S$.

**Contact 8a.** The pitted terrain in Oz Terra (unit $pt_O$) is superposed over the knobbed terrain (*kb*) interpreted to be coeval with unit $la_R$ based on their similar surface textures (i.e., dotted knobs, curvilinear ridges, and the absence of troughs).

**Contact 8b.** The pitted terrain in Vulcan Planitia (unit $pt_V$) is superposed over the smooth-surfaced plains terrain (unit *sp*), which is the oldest terrain of a series of landform units (i.e., *sp*,

*rp*, *la*s, and *la*R) that are emplaced sequentially from south to north over the topographic boundary between Oz Terra and Vulcan Planitia.

**Contact 9.** Lobate-shaped landforms at the base of scarp-bounded ridges and networks of grooves that crosscut the smooth-surfaced plains (unit *sp*) and ridged plains (unit *rp*). Note that northwest-trending linear grooves and ridges (unit *ln*), southward-convex lobate ridges and lobate aprons (unit *lb*), and knobs (unit *k*) are scattered across Vulcan Planitia.

## 5. Discussion

*5.1 Testing Existing Models for the Geomorphologic Development of Charon*

Any models for the tectonic and landscape evolution of Charon must explain the following observations: (1) the absence of ejecta deposits around the largest impact craters (Moore et al., 2016), (2) partial or complete removal of crater rim ridges of the same largest craters (i.e., unit $ct_1$ in Fig. 1C; cf. Fig. 1B) (also see Moore et al., 2016), (3) extensive modification of craters for larger craters mapped as units $ct_1$ and $ct_2$ by superposition of younger landforms across the crater rims and crater floors and erosional removal of crater rim ridges expressed by crosscutting grooves (Figs. 3D, 3E, 4, and 5), (4) truncation and omission of impact craters and crater rim ridges of craters along the eastern edges of the north-trending arcuate ranges in Oz Terra (Figs. 3C, 10A, and 10C), (5) localization of extension along the northern rim of Vulcan Planitia (Moore et al., 2016; Beyer et al., 2017), (6) first-order troughs (i.e., generally > 30 km in width) in Oz Terra that are mutually terminating/crosscutting and orthogonally trending north and east (Figs. 1B and 1C) (Moore et al., 2016), (7) second-order dendritic troughs (generally <~20 km in width) that display hanging valley morphologies at the trough intersections (Figs. 3C, 6A, and 6C), (8) water-ice lobate ridges and lobate aprons (Figs. 3D, 8A, and 8E), (9) parallel and curvilinear round-topped ridges along the northernmost part of Vulcan Planitia (Figs. 3D and 3E), and (10) scattered blocks of various sizes with or without moated margins across Vulcan Planitia.

An ad hoc model may explain one of the above observations. For example, the infilling of crater basins as shown in Figs. 3D-3E, 4 and 5 may be explained by the early proposed cryovolcanism (Moore et al., 2016; Beyer et al., 2019). Although the lack of central- or fissure-style volcanic constructs and the absence of tube-shaped elongated landforms (Moore et al., 2016; Beyer et al., 2019; this study) preclude an extrusive style of cryovolcanism, these observations do not rule out buried cryolava-feeder systems below Vulcan Planitia. Subsurface magmatic plumbing has been inferred for the emplacement of lunar flood basalts (e.g., Andrews-Hanna et al., 2003; Head and Wilson, 2017), which could serve as an analogue for the proposed cryovolcanism in the existing literature (Beyer et al., 2019). Regardless of the style of emplacement, cryovolcanism alone is incapable of explaining the other crater-modification observations such as the absence of impact ejecta deposits and partial or complete removal of rim ridges of the largest craters on Charon's encountered hemisphere (Moore et al., 2016).

The tectonic history of Charon has been inferred from various thermo-mechanical models based on different initial and boundary conditions. In a hot-start model, Rhoden et al. (2015) showed that extensional fractures with longitudinally varying trends can form in the polar and equatorial regions if Charon has an eccentric orbit and a subsurface ocean. The predicted east–west-trending fractures at mid-latitudes by this model are inconsistent with the east-trending extensional belt of Moore et al. (2016) and Beyer et al. (2017) in the equatorial region (Figs. 1B and 1C).

In an alternative hot-start model, the east-trending extensional belt and troughs of various sizes across Oz Terra are interpreted to have been formed by subsurface ocean freezing (Moore et al., 2016; Beyer et al., 2017; Spencer et al., 2021). This interpretation requires an isotropic tensile stress field, which does not explain the localized east-trending extensional belt along the boundary between Oz Terra and Vulcan Planitia. It also does not explain the truncation and omission of craters along the eastern edges of north-trending, eastward-convex arcuate ranges.

Assuming a cold start, Malamud et al. (2017) formulated a one-dimensional thermo-mechanical model that is capable of tracking the evolution of Charon's radius, which in turn predicts its global extension vs. contraction history. Their model considers the effect of compaction of initially porous ice-rock mixtures by self-gravitation, ice-water differentiation through two-phase flow, and heating/cooling due to hydration/dehydration of silicate minerals. Malamud et al. (2017) showed in their model that amorphous ice in the top 10 km of Charon remains unprocessed in the entire history of Charon. Meanwhile, their model predicts early (0-165 Ma) global contraction with the fastest rate at ~140-165 Ma, subsequent global expansion at ~165-450 Ma followed by a constant radius until ~1 Ga, and final global contraction after ~1 Ga. Similar to the ocean-freezing model of (Beyer et al., 2019), the thermo-mechanical model of Malamud et al. (2017) predicts isotropic compressive or extensional stress fields. Such stress states do not explain the localization of the east-striking extensional belt.

Although the lowland region of Vulcan Planitia has been regarded as a site of cryovolcanism (Beyer et al., 2019), we are unable to identify features resembling cryovolcanic constructs such as those interpreted on Titan (Lopes et al., 2007) and Pluto (Schenk et al., 2018b; Cruikshank et al., 2019; Martin and Binzel, 2021; Singer et al. 2022). We are also unable to identify fissure-type eruption centers such as the tiger-stripe fractures on Enceladus (Spencer et al., 2009; Yin and Pappalardo, 2015; Helfenstein and Porco, 2015). Any volcanic conduits responsible for the accumulation of the interpreted cryolavas in Vulcan Planitia must be buried below the surface.

As mentioned above, our mapping does not duplicate the early efforts of crater mapping by Robbins et al. (2017), Singer et al. (2019), and Robbins and Singer (2021). Rather, our mapping focuses on the type of cratering and their relationships to the surrounding landforms. To this end, we identified three types of craters mapped as unit *ct*$_1$, unit *ct*$_2$ and unit *ct*$_3$ in Fig. 1C. Unit *ct*$_1$ representing the oldest craters have their diameters $D>50$ km. They occur exclusively on Oz Terra (see feature 9 in Fig. 1B as an example). Unit *ct*$_2$ is characterized by their sizes D<35 km and their impact deposits displaying rampart-like morphologies as mentioned above (see feature 10 in Fig. 1B as an example) (also see Robbins et al., 2018). Although direct cross-cutting relationships between unit *ct*$_1$ and unit *ct*$_2$ are not observed, the consistently smaller sizes (<35 km) of unit *ct*$_2$ craters and their well-preserved ejecta deposits and rim ridges imply unit *ct*$_2$ to have younger ages. Unit *ct*$_3$ characterized by pristine crater morphology (see feature 11 in Fig. 1B as an example) have their and radial ejecta rays superposed over unit *ct*$_1$ and unit *ct*$_2$ craters. Note that craters of each type are selectively mapped in Fig. 1C to avoid a clogged map presentation of other landform units.

*5.2 Possible Analogues from Earth, Mars, Mercury, and Pluto*

To assist the interpretation of our newly created geomorphological map (Fig. 1C), we systematically compare landforms on Charon with solar-system analogues that have similar shapes and well-understood formation mechanisms. First, dendritic networks of troughs on Charon (Figs. 6A and 6C) resemble glacial valleys (Fig. 6B) or subglacial meltwater channels (Fig. 6D) (Fretwell et al., 2013) on Earth. Note that the trough systems on both solar-system bodies display hanging valleys, steep-walled linear and curvilinear depressions, undulating longitudinal profiles, and overdeepenings (see detailed descriptions in the captions of Figs. 6A-6D). On Earth, undulating longitudinal profiles are characteristic of subglacial meltwater channels (Grau Galofre et al., 2020), and overdeepenings are common along terminal troughs of subglacial meltwater-channel systems (Cook and Swift, 2012). Minor landslides mapped by Beddingfield et al. (2019) are also shown in Fig. 6A. Landslides originated from steeply scarped range flanks in a rift-like valley on Charon resemble those in Valles Marineris (e.g., Lucchitta, 1979) on Mars interpreted as a rift (e.g., Mège and Masson, 1996) or a transtensional left-slip fault zone (Yin, 2012a): the Charon landslides are restricted to the base of the valley-bounding scarps (Beddingfield et al., 2019), whereas the Valles Marineris landslides may extend across most of and in some cases the entire floor of the graben-like valleys (cf., Lucchitta, 1979; Watkins et al., 2015, 2020).

Second, polygonal networks of troughs on Charon (Fig. 6E) are similar in length scales, map patterns, trough-intersection angles, and trough depths to polygonal networks of glacial valleys in the Greenland region on Earth (Fig. 6F). Note that the glaciated valleys on Earth are exposed in the foreland of the active Greenland sheet that was much larger and covered the entire land and shelf areas of Greenland during the last glacial maximum (Funder et al., 2011). Hence, the region shown in Fig. 5F could have been generated initially below an ice sheet followed by valley glaciation.

Third, the troughed highland terrain (unit *th* in Fig. 1C) and the knobbed plains (unit *sp*) along the northern margin of Vulcan Planitia (Fig. 7A) resemble those on Mars created by rock glaciers (Fig. 7B) (Kronberg et al., 2007). Both areas host partially filled craters, U-shaped valleys, and scattered knobs, rampart craters, and moated mounds on the foreland plains. In addition, both areas expose lobate aprons and knob trains (Figs. 7C and 7D). The latter is interpreted to have been created by rock glaciers based on comparisons against analogues from the Earth (Kronberg et al., 2007). The most direct comparison of lobate aprons on Charon (Fig. 7C) is those exposed on Pluto

(Fig. 7E), which are interpreted to have been induced by $N_2$-ice glaciation with water-ice boulders as glacial tills forming the lateral and frontal moraines (Howard et al., 2017). Conducting more detailed comparison of landforms between Pluto and Charon is limited by the resolutions of the images. However, the occurrence of lobate-shaped landforms on both bodies are undeniable; their close association with isolated knobs or knob trains is readily explained as glacial moraines as discussed in Howard et al. (2017). This in turn implies that Charon may have experienced $N_2$ glaciation. Because of the surface of Charon is composed of ammoniated water ice that is involatile (Protopapa et al., 2021), the inferred glaciers on Charon may have been composed of volatile $N_2$ ice that was removed during deglaciation (i.e., much like water ice on Earth) and involatile ammoniated $H_2O$ ice (much like the rock debris on Earth) that stayed after deglaciation.

Fourth, the ridged plains on Charon (Fig.8A) resemble piedmont moraine ridges on Earth (Fig. 8B) (Sharp, 1958). In this comparison, the curvilinear round-topped ridges on Charon represent folds, whereas the isolated blocks mingled with the ridges may represent erratic boulders.

Fifth, ridge flanks with linear grooves and linear trains of smaller knobs on Charon (feature 1 in Fig. 8C) resemble a glacially striated surface on a ridge flank on Earth (Fig. 8D) (Jakobsson et al., 2016). Parallel sets of evenly spaced linear grooves and ridges on glaciated surfaces are generally referred to as glacial flutes (Gordon et al., 1992). They are known as mega-scale glacial lineations when the length exceeds >5-10 km on Earth (Clark, 1993). The linear grooves and linear knob chains on the ridge flank shown in Fig. 8C are >30-50 km in length, which qualify them as mega-scale glacial lineations should they have occurred on Earth. Mega-scale glacial lineations on Earth are created by fast-flowing ice streams, which are like rivers, are bounded by stagnant ice on the two sides (Clark, 1993; King et al., 2009). Mega-scale glacial lineations up to ~17 km in length hosted by a glaciated landscape have been documented in the Tharsis rise region of Mars (Yin et al., 2021).

Sixth, lobate ridges on Charon (Fig. 8E) resemble those created by rock glaciers on Earth (Elliott-Fisk, 1987) (Fig. 8F). In this comparison, the convex direction of the lobate ridges would indicate the glacial flow direction, whereas boulders associated with the lobate ridges on Charon would represent glacial erratic deposits.

Seventh, moated mounds and rampart craters on Charon and Mars (Kronberg et al., 2007) are similar in shapes but differ by their topographic reliefs: the relief on Charon is ~3 km (Figs. 9A and 9B) whereas the relief on Mars is only ~100 m (Figs. 9C and 9D). The general similarities of the rampart craters on Charon and Mars were noted earlier (Kronberg et al., 2007); both display higher rim ridges and lower ejecta-terminus ridges as shown in Figs. 9E-9H. The origin of rampart craters on Mars has been attributed to the presence of ground ice or buried glaciers (Carr et al., 1977; Weiss and Head, 2013). For the moated mounds acting as rock debris flowing on top of a viscous glacial material, their elevations above the flat surrounding surface should have been controlled by isostasy: higher mounds with deeper roots and lower mounds with shorter roots.

Finally, deformation-modified craters on Charon are best expressed by crater-rim offset, removal, and truncation, which are closely associated with linear or curvilinear north-trending scarps that bound the north-trending ranges (Figs. 10A and 10C). The crater-offset scarps in Fig. 10A and those shown in Fig. 3B (see detailed description in its caption) resemble the thrust-induced scarps on Mercury (Fig. 10B) (Watters et al., 2015). The truncated craters on Charon as shown in Figs. 10A and 10C resemble the crater-truncated Thaumasia thrust (Fig. 10D) bounding the eastern edge of the Tharsis rise on Mars (Nam and Schultz, 2010; Yin, 2012b).

*5.3 A Landscape Evolution Model of Charon*

To account the major observations listed above, a system approach must be adopted. Specifically, the system-based landscape model must explain the spatiotemporal relationships of all landform units and major landform features established by the early studies (Moore et al., 2016; Beyer et al., 2017, 2019; Robbins et al., 2017, 2018, 2019) and this study (Fig. 1). Here, we propose a self-consistent tectono-geomorphological model for the geological evolution of Charon's encounter hemisphere (Fig. 11). The model shown in Fig. 11 is inspired by the well-established glacial landsystem-evolution models on Earth (e.g., Brodzikowski and van Loon, 1987) that predict many landforms with similar shapes as we mapped. The key ingredients of the model include (1) the early formation of the impact-induced east-trending extensional belt along the northern margin of Vulcan Planitia, (2) younger north-trending thrusting due to despinning of Charon, (3) coeval thrusting and glaciation in the highland region (Oz Terra), and (4) glacial transport of water-ice debris into the Vulcan Planitia impact basin. The main supports of past glaciation on Charon come from their similar shapes to the landforms on Earth, Mars, and Pluto that have known glacial/ground-ice origins (Figs. 7 and 9). The support for east-west compression that created the north-trending, eastward convex, arcuate ranges come from comparison of truncated and omission of impact craters induced by thrusting on Mercury and Mars (Figs. 10A-10D).

In our model, we envision that Charon experienced an initial phase of intense bombardments, expressed by the occurrence of largest impact craters (D>50-100 km) exposed on the countered hemisphere. Following the early suggestion of Malamud et al. (2017), a giant impact created Vulcan Planitia that was much deeper than its present depth (Fig. 11A). The ridge-valley zone (unit *rv* in Fig. 1C) along the northern rim of the interpreted impact basin is interpreted as an extensional fault zone created by the giant impact (Fig. 11B). Note that Malamud et al. (2017) considered the east-trending extensional belt to have been created by global contraction followed by extensional reactivation, which is different from the scenario proposed here. An interesting consequence of the giant impact hypothesis is its injection of heat into Charon, and the elevated temperatures would have favored not only cryovolcanism as suggested by Malamud et al. (2017) and the formation of a transient $N_2$-dominated atmosphere capable of driving $N_2$ glaciation.

A major issue with the giant impact hypothesis of Malamud et al. (2017) is the lack of impact ejecta deposits in Oz Terra (Beyer et al., 2019). In our model, we suggest that $N_2$-ice glaciation in Oz Terra, which occurred after the giant impact, removed the impact-breccia deposits and transported them to the Vulcan Planitia basin. The same glacial-erosion process was also responsible for the complete removal of impact ejecta deposits around the largest craters in Oz Terra and partial or complete removal of their crater rim ridges.

The infilling of Vulcan Planitia by cryovolcanism could have occurred as envisioned by Beyer et al. (2019) prior to or during our proposed glaciation. However, cryovolcanism must have terminated prior to the final emplacement of water-ice-debris-bearing $N_2$ glaciers, which are expressed as parallel round-topped ridges mingled with blocks of various sizes up to 10-20 km and shapes (Fig. 11B). The ridges are interpreted as water-ice moraine landforms and blocks as water-ice erratic boulders (Fig. 11B).

Supported by the similar patterns of trough networks on Charon and a glaciated landform on Earth (Figs. 6E and 6F), we suggest that glacial erosion was responsible for creating the orthogonal network of the first-order troughs in Oz Terra (Fig. 11C). These channels could have been developed either below an ice sheet, or as glacial transport U-shaped valleys as on Earth shown in Fig. 6F.

Similar morphologies for the truncated and omitted craters on Charon, Mercury, and Mars (Figs 10A-10D) led us to suggest that the north-trending arcuate ranges were created by thrusting during east-west compression. In this interpretation, the eastward convex-shaped ranges should have been created by west-dipping thrusts. We suggest that the apparent offset of the north-trending ranges (features 1 and 2 in Fig. 1B) on the two sides of the valley zone (unit *rv* in Fig. 1C) was a result of crack arresting (Cooke and Underwood, 2001); that is, the older and weaker faults along the eastern extensional belt were reused as strike-slip transfer structures linking north-trending thrusts on its two sides (Figs. 11D and 11E). The lack of rift shoulders along the western segment of the valley zone (i.e., unit *rv* in Fig. 1C) can be explained by the syn- and post-thrusting glacial erosion (Figs. 11D-11E).

Post-thrusting glaciation is not only expressed by the removal of rift-shoulder ridges but also by the development of younger dendritic networks of troughs (i.e., unit $tr_2$ in Fig. 1C) (Fig. 11E). Possible Earth analogues (Figs. 6B and 6D) imply that the dendritic troughs on Charon could have been formed either by valley glaciers or subglacial melt-channel development. The valley-glacier interpretation implies cold-based glaciation (i.e., the basal ice is frozen with the underlying bedrock and glacial movement is accommodated by ice creeping within the glaciers), while subglacial meltwater-channel interpretation implies at least local warm-based glaciation (i.e., melting occurs at the base of an ice sheet). Under the cold-based condition, the pre-glacial landforms could be preserved as glacial flows are mainly accommodated by ice-sheet deformation without the involvement of subglacial tills and subglacial bedrock (e.g., Waller, 2001). In contrast, wet-based glaciation should have created streamlined landforms (e.g., Evans et al., 2006), which are absent next to the interpreted dendritic channels. This led us to suggest that the dendritic channels may have been developed as glacial valleys, not meltwater conduits. This interpretation is consistent with the presence of hanging valleys in the dendritic channel systems (e.g., Fig. 3C). Some dendritic channels on Charon are associated with pyramids, arêtes, U-shaped valleys, and hanging valleys (see Figs. 5E and 5F), which are best explained by their formation of valley glaciation.

Based on the analogues from Mars (Kronberg et al., 2007) (Figs. 7B-7D and 9C), we speculate that Vulcan Planitia may have been filled by rock glaciers; that is, water ice as the bedrock debris that covers $N_2$ ice below (Fig. 11E). In this interpretation, the polygonal-shaped blocks/mounds/mountains of various sizes (up to >30 km across) with or without moated margins are interpreted as floating water-ice blocks. Their various elevations are controlled by isostasy: higher-elevation blocks, mounds, and mountains in Vulcan Planitia should have deeper roots extending into the low-viscous nitrogen ice below.

We suggest that the formation of recessional moraines resulted in the development of lobate ridges, lobate aprons, knob trains, and isolated knobs composed of water ice. The isolated knobs in Oz Terra are interpreted as to be the same as the blocks in Vulcan Planitia, which represent erratic boulders composed of water ice and originated from impact breccia (Fig. 11F).

It is likely that deglaciation was associated with rapid sublimation and eventual removal of $N_2$ ice under a thinner atmosphere. The cooling due to atmospheric thinning may have created tensile fractures across Vulcan Planitia, resulting from unstable extension of a rheologically stratified system (Fletcher and Hallet, 1983; Yin, 2000); in this case, the brittle water-ice debris layer overlies the viscous $N_2$-ice layer. The rapid cooling event may have also frozen mega-breccia of water ice on the way to sink into the subsurface viscous $N_2$-ice layer below Vulcan Planitia. The rapid freezing is expressed as moated margins around the polygonal mounds/mountains and the formation of funnel-like depressions (Fig. 1C).

The thickness of glaciers across Oz Terra is estimated to ~2-3 km, which is constrained by the topographic relief of the first-order orthogonal troughs interpreted as glacial valleys (i.e., unit *tr₁* in Fig. 1C; also see Figs. 2A-2D). In contrast, the thickness of the glaciers across the east-trending valley zone (unit *rv* in Fig. 1C) along the boundary between Oz Terra and Vulcan Planitia is estimated to be >~5 km, which is constrained by the elevation difference between the striated ridge flank that is interpreted as the site of ice-stream flow and its bounding valley floors interpreted as regions of stagnant ice (see Fig. 8C and related interpretations).

*5.4 Plausibility of our Proposed Glaciation Model*

The similar landforms on Charon (this study) and Pluto (Howard et al., 2017) to those created by glaciation on Earth raise the question of why these solar-system bodies with drastically different gravities, surface compositions, and atmospheres would have operated in such a way. As argued by Collins et al. (2010), the similar homologous temperature ($T_H$) of rocks on Earth and water ice in the outer solar system may have played a first-order role in generating similar styles and shapes of tectonic structures on Earth and icy satellites. The homologous temperature of a material is defined by the ratio of the averaged deformation temperature ($T_D$) and its melting temperature ($T_M$) (Goetze, 1978; Weertman, 1983). Below, we compare the homologous temperatures of glaciers and bedrock on Earth and Charon

In the polar regions of the Earth, $T_H$ for $H_2O$ ice is ~0.81 assuming $T_D$ = 223 K and $T_M$ = 273 K, and $T_H$ for felsic rocks is ~0.24 assuming $T_D$ = 223 K and $T_M$ = 923 K. In comparison, glaciers on Charon are likely composed of $N_2$ ice (although they have been sublimated away by now), similar to the interpreted $N_2$-ice glaciers carrying water-ice debris on Pluto (Howard et al., 2017). The bedrock of Charon is composed of ammoniated water ice ($NH_3 \cdot nH_2O$) (Protopapa et al., 2021) that has $T_D$ = ~50 K, $T_M$ (N$_2$) = ~63 K, and $T_M$ (NH$_3$•nH$_2$O) = ~176 K, respectively; these values yield $T_H$ for N$_2$ ice on Charon to be ~0.79 (cf. 0.81 for water ice on Earth) and $T_H$ for NH$_3$•nH$_2$O ice on Charon to be ~0.28 (cf. 0.24 for felsic rocks on Earth). The above comparison indicates that the homologous temperatures of glaciers and bedrocks on Earth and Charon are nearly identical.

In order to explain the ~30:1 ratio for the maximum heights of the moated mounds on Charon (~3 km) and Mars (~100 m), we assume the inferred bedrock debris on top of viscous glaciers to have plastic strengths of $Y_C$ and $Y_M$ for Charon and Mars, respectively. Equating the plastic strength to the vertical load requires $Y_C = \rho_C g_C h_C$ and $Y_M = \rho_M g_M h_M$, where $\rho_C$ = 1000 kg m$^{-3}$ is the water ice density, $\rho_M$ = 2300 kg m$^{-3}$ is rock debris density, and $g_C$ and $g_M$ are the surface gravity of Charon (0.288 m s$^{-2}$) and Mars (3.72 m s$^{-2}$). Because brittle materials such as rocks and water ice have similar mechanical strengths (Byerlee, 1978; Schulson, 2001), $Y_C/Y_M \approx$ 1 would require $h_C/h_M \approx 29.7$ using the physical quantities mentioned above. This estimated $h_C/h_M$ ratio is very similar to the observed ratio of ~30:1. The above analysis implies that the debris-cover layer is a viscoplastic material: a block sinks if its weight-induced pressure is larger than the plastic yield strength of the layer but stays on top if its weight-induced pressure is smaller than the yield strength. Note that the elevation of the blocks, once sunk into the subsurface viscous layer of nitrogen ice, are controlled by isostasy.

*5.5 Predictions of the Proposed Landscape Model*

Our proposed landscape-evolution model (Fig. 11) provides new insights into some of the interesting observations made by previous researchers. The lower-than-expected number of D≲10-20 km craters in Vulcan Planitia (Singer et al., 2019) can be accounted for by a combination of glacial erosion, glacial deposition, and glacier flow possibly superposed on top of an earlier fluvial landscape. This means that the size-frequency distribution (SFD) of craters on Charon should be regarded only as a lower bound when used to test competing models of planet formation during the evolution of the Solar System (e.g., Schlichting, 2011). Additionally, the landscape model in Fig. 11 provides an alternative explanation of craters on Charon that are consistently shallower than those of Saturn's icy moons with similar surface gravity and ice compositions (Schenk et al., 2018a). Schenk et al. (2018a) suggest that the shallower crater depths on Charon may have been induced by slower impact velocities in the Kuiper belt. In our model, we suggest that the shallower depth can be demonstrated as a combined result of rim-ridge erosion and crater-basin infilling.

As shown by Malamud et al. (2017), amorphous water ice in their cold-start model should have been un-processed by heating throughout the evolution of Charon. If Charon was evolved in such a way, its surface exposure of crystalline water ice exclusively along the north-trending ranges in Oz Terra requires an explanation (Protopapa et al., 2021). In our model, the exposure of the crystalline ice in these mountain ranges is explained by thrusting and glacial erosion. That is, deeper crystallized water ice formed at hotter temperatures were brought up to the surface by thrusting. The alternative hot-start models (e.g., Beyer et al., 2019) would predict global exposure of crystalline water ice. But such a model does not explicitly explain where and why the highest volumes and largest grain sizes of crystalline water ice are exposed only along the north-trending ranges in the encounter hemisphere of Charon (Protopapa et al., 2021).

Our interpreted north-trending thrusting as the cause of arcuate-range formation in the equatorial and low latitude regions is compatible with a stress field generated by despinning of Charon during its receding course from but before being tidally locked on a circular orbit around Pluto (Matsuyama and Nimmo, 2013; Barr and Collins, 2015; Rhoden et al., 2020). However, despinning alone does not explain why the range-bounding thrusts dip consistently to the west, as required by the eastward convex shape of the arcuate ranges. On Earth, thrusts with the same dip direction are commonly parts of a structural system bounded below by a unidirectionally moving basal decollement (e.g., Boyer and Elliott, 1982). Based on this insight, we interpret the west-dipping thrusts on Charon to be parts of an east-directed thrust system (Fig. 12). We suggest that the mechanically favored location for the inferred decollement may lie along the brittle-ductile transition zone and the top-east flow in the ductile ice may have been driven by warm-ice convection. This speculation could be further modelled in future studies.

The timing and duration of our proposed glaciation are unconstrained. Based on the current dating using crater size-frequency distributions (Singer et al., 2019), our mapped glaciated terrains should have ages of >4.0 Ga. This means that the maximum duration of glaciation could be as long as ~500 Ma. Such a protracted event is possible considering the recent work on Pluto's atmosphere that shows its escape rate is four orders of magnitude slower than the theoretical expectation (Gladstone et al., 2016), possibly due to the presence of dust in Pluto's atmosphere (Zhang et al., 2017; Lavvas et al., 2021). Glaciation on Charon could have occurred as a single event, or as multiple events. The latter may have been induced by Charon's complex climate history (Hansen and Paige, 1996; Bertrand et al., 2018; Johnson et al., 2021) or comet impact events (Stern et al., 2015).

Our proposed landscape-evolution model makes the following testable predictions. First, the knobs scattered in Vulcan Planitia should have different volume contents and grain sizes of crystalline water ice, reflecting their different origins at various depths of the ice shell in Oz Terra excavated to the surface by impacts and transported to the lowland plains by glaciation. This prediction can be tested by using spatially resolved IR spectra acquired with the Linear Etalon

Imaging Spectral Array (LEISA), which is part of the New Horizons Ralph instrument (Reuter et al., 2008). Because the best LEISA scan of Charon is at a spatial resolution of ~5 km/pixel (Protopapa et al., 2021), only the largest knobs can be differentiated by this data set. Second, Vulcan Planitia should contain debris-covered glaciers. This prediction can be tested using orbitally based ground-penetrating radar techniques that were used to have identified rock glaciers on Mars (Holt et al., 2008). Third, the inferred thrusts should have generated foreland-basin deposits, and their internal beds should be thickening towards the interpreted thrust traces and the beds next to the thrusts should be folded. Similarly, the western segment of the valley zone interpreted to have an extensional origin should have preserved rift-basin deposits although the valley-bounding rift shoulders have been eroded away. The rift-basin deposits should show tilted strata dipping opposite to the dip of the nearby basin-bounding fault. The above predictions between faults and syn-faulting sedimentation can be tested using orbitally based ground-penetrating radar techniques (Phillips et al., 2008). Fourth, Vulcan Planitia as a giant impact basin can be tested when complete imaging of Charon is achieved.

*5.6 Implications for the Evolution of Larger KBOs*

An implication of our work is that larger KBOs with sizes equal to or greater than Charon may have evolved from an initial stage (Figs. 13A and 13D) characterized by the presence of a $N_2$-dominated atmosphere capable of driving $N_2$-liquid activities. A possible analogue for this stage of evolution is Titan but with an important difference: the fluvial agent on Charon is nitrogen whereas the fluvial agents on Titan are methane and ethane (Stofan et al., 2007; Lorenz et al., 2008; Hayes et al., 2018). The gravity and atmospheric pressure conditions may also differ between the two solar-system bodies. Here, we simply emphasize the possible surface processes during this initial stage on larger KBOs. Although there is no direct evidence for fluvial activities on Charon, it is possible that the interpreted dendritic glacial-valley networks originated from fluvial channels. The duration of this stage is unconstrained, but most likely occurred in the first few million years after the Charon-forming impact with a temperature enhancement of ~30 K (McKinnon et al., 2021) or immediately after the giant impact that created the Vulcan Planitia basin.

During the second stage (Figs. 13B and 13D), Jeans and hydrodynamic escape in conjunction with impact erosion may have caused the gradual thinning of the primordial atmosphere of the larger KBO. Meanwhile, termination of gravitational accretion and waning impact bombardments reduced surface heating. The combined effect of atmospheric thinning and reduction of heat release have led to surface cooling, which allowed the transition from a landscape shaped mainly by fluvial processes to a landscape shaped mainly by glaciation. Pluto with a thin atmosphere and ongoing glaciation may serve as a modern analogue for this stage (Fig. 13B). The duration for this stage may be as long as up to 500 Ma as exemplified by Charon.

In the final stage, the primordial atmosphere was completely removed and the internal heat sources were largely exhausted. The resulting low surface pressure caused rapid sublimation of $N_2$-ice glaciers, which in turn caused rapid cooling and the runaway $N2$-ice removal. This process must have occurred rapidly so that viscous-deformation processes expressed by moated mounds and funnel-like depressions were frozen into the modern landscape (Figs. 13C and 13D). Current Charon without a detectable atmosphere may serve as an example of this stage.

The time scale for the rapid cooling phase that led to the creation of the moated mountains and funnel-like depressions should be less than ~1000 years based on a simple calculation below. We assume that the moated mounds in Vulcan Planitia are sinking rigid blocks and their average density is ~1 kg/m$^3$ greater than the underlying N$_2$ ice. For a spherical mound with a diameter of ~20 km, the downward velocity of the mound according to the Stokes law would be ~10$^{-10}$ m/s given the N$_2$-ice viscosity of ~10$^{14}$ Pa s. Using the moat depth of 1-2 km (Fig. 9) as a constraint for the magnitude of the downward motion, the time scale required for the sinking mound is between 300 and 600 years. Note that the above estimates depend critically on the viscosity of N$_2$ ice. The lack of information on the strain rate during the inferred sinking of the moated mounds, a critical factor in determining the effective viscosity of N$_2$ ice (Yamashita et al., 2010), prevents us from exploring this issue any further.

**Conclusions**

A systematic geomorphological mapping and detailed landform analysis using the highest resolution images obtained by the New Horizons spacecraft reveal the presence of a range of differentiable terrains that were not examined in detailed by the early studies. These terrains and their cross-cutting relationships define a complex spatiotemporal evolution of the ice shell and the surface morphology of Charon's encountered hemisphere. The most important findings of our work include:

(1) truncation and omission of large craters (diameters > 30-40 km) and their crater rim ridges along the eastern edges of several north-trending, eastward-convex, arcuate ranges in Oz Terra of the northern encountered hemisphere;

(2) lobate ridges, lobate knob trains, and lobate aprons resembling glacial moraine landforms on Earth;

(3) dendritic channel systems containing hanging valleys; and

(4) locally striated surfaces defined by parallel ridges, knob trains, and grooves that are >40-50 km in length.

The above observations and the topographic dichotomy of Charon's encountered hemisphere can be explained by a landscape-evolution model that involves (i) a giant impact that created the Vulcan Planitia basin and the extensional fault zone along its northern rim, (ii) a transient atmosphere capable of driving $N_2$-ice glaciation eroding the water-ice bedrock and transporting water-ice debris to sedimentary basins, (iii) regional glacial erosion and transport of earlier emplaced impact eject deposits from the highlands of Oz Terra into the lowland basin of Vulcan Planitia, (iv) syn-glaciation north-trending thrusting interpreted to have been induced by Charon's despinning, and (v) the development of a water-ice debris cover layer over subsurface $N_2$ ice below Vulcan Planitia during global deglaciation. The infilling of the Vulcan Planitia could have been accompanied by cryovolcanism. The extensive modification of impact craters means that the crater size-frequency distributions from Charon should serve only as a lower bound when used to test the formation mechanism of Kuiper belt objects.

**Acknowledgements**. This research would have never been possible without the dedicated effort of the New Horizons Team, whose persistence and intellectual leadership led to the exciting discoveries of a remote world that was unknown to humans. We thank the New Horizons Team especially for making data not only available to the public but in a format for outsiders to use directly for meaningful research. This study is a result of curiosity-driven research not funded by any agencies. Our work benefited greatly from discussions with our colleagues Dave Paige and Hilke Schlichting at UCLA.

**References**

Arakawa, S., R. Hyodo, H. Genda, Early formation of moons around large trans-Neptunian objects via giant impacts. *Nature Astronomy*, **3**(9), 802-807 (2019).

Abod, C. P., Simon, J. B., Li, R., Armitage, P. J., Youdin, A. N., & Kretke, K. A. (2019). The mass and size distribution of planetesimals formed by the streaming instability. II. The effect of the radial gas pressure gradient. The Astrophysical Journal, 883(2), 192.

Barr, A. C., G. C. Collins, Tectonic activity on Pluto after the Charon-forming impact. *Icarus*, **246**, 146-155 (2015).

Beddingfield, C. B., R. Beyer, W. B. McKinnon, K. Runyon, W. Grundy, S. A. Stern, V. Bray, R. Dhingra, J. M. Moore, K. Ennico, C. B. Olkin, P. Schenk, J. R. Spencer, H. A. Weaver, L. A. Young, and the New Horizons Team, Landslides on Charon, *Icarus*, **335**, 113383 (2019).

Bertrand, T., F. Forget, O. M. Umurhan, W. M. Grundy, B. Schmitt, S. Protopapa, A. M. Zangari, O. L. White, P. M. Schenk, K. N. Singer, A. Stern, H. A. Weaver, L. A. Young, K. Ennico, C. B. Olkin, The nitrogen cycles on Pluto over seasonal and astronomical timescales. *Icarus*, **309**, 277-296 (2018).

Beyer, R. A., F. Nimmo, W. B. McKinnon, J. M. Moore, R. P. Binzel, J. W. Conrad, A. Cheng, S. J. Desch, M. Neveu, Differentiation and cryovolcanism on Charon: A view before and after New Horizons. *Icarus*, **287**, 175-186 (2017).

Beyer, R. A., J. R. Spencer, W. B. McKinnon, F. Nimmo, C. Beddingfield, W. M. Grundy, K. Ennico, J. T. Keane, J. M. Moore, C. B. Olkin, S. Robbins, K. Runyon, P. Schenk, K. Singer, S. A. Stern, H. A. Weaver, L. A. Young, the New Horizons Team, The nature and origin of Charon's smooth plains. *Icarus*, **323**, 16-32 (2019).

Bierson, C. J., F. Nimmo, W. B. McKinnon, Implications of the observed Pluto–Charon density contrast. *Icarus*, **309**, 207-219 (2018).

Bland, M. T., K. N. Singer, W. B. McKinnon, P. M. Schenk, Enceladus' extreme heat flux as revealed by its relaxed craters. *Geophysical Research Letters*, **39**(17), L17204 (2012).

Boyer, S. E., and D. Elliott, Thrust systems. *American Association of Petroleum Geology Bulletin*, **66**(9), 1196-1230 (1982).

Brodzikowski, K., A. J. Van Loon, A systematic classification of glacial and periglacial environments, facies and deposits. *Earth-Science Reviews*, **24**(5), 297-381 (1987).

Byerlee, J., Friction of rocks, in *Rock Friction and Earthquake Prediction* (Birkhäuser, Basel, 1978), pp. 615-626.

Canup, R. M., A giant impact origin of Pluto-Charon. *Science*, **307**(5709), 546-550 (2005).

Canup, R. M., K. M. Kratter, M. Neveu, "On the origin of the Pluto system" in *The Pluto System After New Horizons* (Univ. of Arizona, Tucson, 2021), pp. 475–506.

Carr, M. H., L. S. Crumpler, J. A. Cutts, R. Greeley, J. E. Guest, H. Masursky, Martian impact craters and emplacement of ejecta by surface flow. *Journal of Geophysical Research*, **82**(28), 4055-4065 (1977).

Cheng, A. F., H. A. Weaver, S. J. Conard, M. F. Morgan, O. Barnouin-Jha, J. D. Boldt, K. A. Cooper, E. H. Darlington, M. P. Grey, J. R. Hayes, K. E. Kosakowski, T. Magee, E. Rossano, D. Sampath, C. Schlemm, H. W. Taylor, Long-range reconnaissance imager on New Horizons. *Space Science Reviews*, **140**(1–4), 189–215 (2008).

Collins, G. C., W. B. McKinnon, J. M. Moore, F., Nimmo, R. T. Pappalardo, L. M., Prockter, P. M. Schenk, Tectonics of the outer planet satellites. *Planetary Tectonics*, **11**(264), 229 (2010).

Conrad, J. W., F. Nimmo, R. A. Beyer, C. J. Bierson, P. M. Schenk, Heat flux constraints from variance spectra of Pluto and Charon using limb profile topography. *Journal of Geophysical Research: Planets*, **126**(2), e2020JE006641 (2021).

Cook, S. J., D. A. Swift, Subglacial basins: Their origin and importance in glacial systems and landscapes. *Earth-Science Reviews*, **115**(4), 332-372 (2012).

Cooke, M. L., C. A. Underwood, Fracture termination and step-over at bedding interfaces due to frictional slip and interface opening. *Journal of structural geology*, **23**(2-3), 223-238 (2001).

Cruikshank, D. P., O. M. Umurhan, R. A. Beyer, B. Schmitt, J. T. Keane, K. D. Runyon, D. Atri, O. L. White, I. Matsuyama, J. M. Moore, W. B. McKinnon, S. A. Sandford, K. N. Singer, W. M. Grundy, C. M. Dalle Ore, J. C. Coo, T. Bertrand, S. A. Stern, C. B. Olkin, H. A. Weaver, L. A. Young, J. R. Spencer, C. M. Lisse, R. P. Binzel, A. M. Earle, S. J. Robbins, G. R. Gladstone, R. J. Cartwright, K. Ennico, Recent cryovolcanism in virgil fossae on Pluto. *Icarus*, **330**, 155-168 (2019).

Elliott-Fisk, D. L., Glacial geomorphology of the White Mountains, California and Nevada: Establishment of a glacial chronology. *Physical Geography*, **8**(4), 299-323 (1987).

Ennico, K., T. R. Lauer, C. B. Olkin, S. Robbins, P. Schenk, K. Singer, J. R. Spencer, S. A. Stern, H. A. Weaver, L. A. Young, A. M. Zangari, Charon tectonics. *Icarus*, **287**, 161-174 (2017).

Evans, D. J. A., E. R. Phillips, J. F. Hiemstra, C. A. Auton. Subglacial till: formation, sedimentary characteristics and classification. *Earth-Science Reviews*, **78**, 115-176 (2006).

Farinella, P., & Davis, D. R. (1996). Short-period comets: primordial bodies or collisional fragments? Science, **273**(5277), 938-941.

Fergason, R. L., T. M. Hare, J. Laura, HRSC and MOLA blended digital elevation model at 200m v2. Astrogeology PDS Annex, US Geological Survey (2018).

Fletcher, R. C., B. Hallet, Unstable extension of the lithosphere: A mechanical model for basin-and-range structure. *Journal of Geophysical Research: Solid Earth*, **88**(B9), 7457-7466 (1983).

Fraser, W.C., M. E. Brown, A. Morbidelli, A. Parker, K. Batygin, The absolute magnitude distribution of Kuiper belt objects. *The Astrophysical Journal*, **782**(2), 100 (2014).

Fray, N., Schmitt, B., Sublimation of ices of astrophysical interest: A bibliographic review. *Planetary and Space Science*, **57**, 2053-2080 (2009).

Fretwell, P., H. D. Pritchard, D. G. Vaughan, J. L. Bamber, N. E. Barrand, R. Bell, C. Bianchi, R. G. Bingham, D. D. Blankenship, G. Casassa, G. Catania, D. Callens, H. Conway, A. J. Cook, H. F. J. Corr, D. Damaske, V. Damm, F. Ferraccioli, R. Forsberg, S. Fujita, P. Gogineni, J. A. Griggs, R. C. A. Hindmarsh, P. Holmlund, J. W. Holt, R. W. Jacobel, A. Jenkins, W. Jokat, T. Jordan, E. C. King, J. Kohler, W. Krabill, M. Riger-Kusk, K. A. Langley, G. Leitchenkov, C. Leuschen, B. P. Luyendyk, K. Matsuoka, Y. Nogi, O. A. Nost, S. V. Popov, E. Rignot, D. M. Rippin, A. Riviera, J. Roberts, N. Ross, M. J. Siegert, A. M. Smith, D. Steinhage, M. Studinger, B. Sun, B. K. Tinto, B. C. Welch, D. A. Young, C. Xiangbin, A. Zirizzotti, Bedmap2: improved ice bed, surface and thickness datasets for Antarctica. *The Cryosphere*, **7**(1), 375-393 (2013).

Funder, S., K. K. Kjeldsen, K. H. Kjær, C. Ó. Cofaigh, The Greenland Ice Sheet during the past 300,000 years: A review. *Developments in Quaternary Sciences*, **15**, 699-713 (2011).

Gladstone, G. R., S. A. Stern, K. Ennico, C. B. Olkin, H. A. Weaver, L. A. Young, M. E. Summers, D. F. Strobel, D. P. Hinson, J. A. Kammer, A. H. Parker, A. J. Steffl, I. R. Linscott, J. Wm. Parker, A. F. Cheng, D. C. Slater, M. H. Versteeg, T. K. Greathouse, K. D. Retherford, H. Throop, N. J. Cunningham, W. W. Woods, K. N. Singer, C. C. C. Tsang, C. M. Lisse, M. L. Wong, Y. L. Yung, X. Zhu, W. Curdt, P. Lavvas, E. F. Young, G. L. Tyler, the New Horizons

Science Team, The atmosphere of Pluto as observed by New Horizons. *Science*, **351**(6279), aad8866 (2016).

Goetze, C. The mechanisms of creep in olivine. *Philosophical Transactions of the Royal Society of London. Series A, Mathematical and Physical Sciences*, **288**(1350), 99-119 (1978).

Grau Galofre, A. A. M. Jellinek, G. R. Osinski, Valley formation on early Mars by subglacial and fluvial erosion. *Nature Geoscience*, **13**(10), 663-668 (2020).

Hansen, C. J., D. A. Paige, Seasonal nitrogen cycles on Pluto. *Icarus*, **120**(2), 247-265 (1996).

Hayes, A. G., R. D. Lorenz, J. I. Lunine, A post-Cassini view of Titan's methane-based hydrologic cycle. *Nature Geoscience*, **11**(5), 306-313 (2018).

Helfenstein, P., C. C. Porco, Enceladus' geysers: Relations to geological features. *The Astronomical Journal*, **150**(3), 96 (2015).

Holt, J. W., A. Safaeinili, J. J. Plaut, J. W. Head, R. J. Phillips, R. Seu, S. D. Kempf, P. Choudhary, D. A. Young, N. E. Putzig, D. Biccari, Y. Gim, Radar sounding evidence for buried glaciers in the southern mid-latitudes of Mars. *Science*, **322**(5905), 1235-1238 (2008).

Howard, A. D., J. M. Moore, O. M. Umurhan, O. L.White, R. S. Anderson, W. B. McKinnon, J. R. Spencer, P. M. Schenk, R. A. Beyer, S. A. Stern, K. Ennico, C. B. Olkin, H. A. Weaver, L. A. Young, the New Horizons Science Team, Present and past glaciation on Pluto. *Icarus*, **287**, 287-300 (2017).

Howett, C. J. A., A. H. Parker, C. B. Olkin, D. C. Reuter, K. Ennico, W. M. Grundy, A. L. Graps, K. P. Harrison, H. B. Throop, M. W. Buie, J. R. Lovering, S. B. Porter, H. A. Weaver, L. A. Young, S. A. Stern, R. A. Beyer, R. P. Binzel, B. J. Buratti, A. F. Cheng, J. C. Cook, D. P. Cruikshank, C. M. Dalle Ore, A. M. Earle, D. E. Jennings, I. R. Linscott, A. W. Lunsford, J. Wm. Parker, S. Phillippe, S. Protopapa, E. Quirico, P. M. Schenk, B. Schmitt, K. N.

Singer, J. R. Spencer, J. A. Stansberry, C. C. C. Tsang, G. E. WeigleII, A. J. Verbiscer, Inflight radiometric calibration of New Horizons' multispectral visible imaging camera (MVIC). *Icarus*, **287**, 140–151 (2017).

Jakobsson, M., J. Nilsson, L. Anderson, J. Backman, G. Björk, T. M. Cronin, N. Kirchner, A. Koshurnikov, L. Mayer, R. Noormets, M. O'Regan, C. Stranne, R. Ananiev, N. B. Macho, D. Cherniykh, H. Coxall, B. Eriksson, T. Flodén, L. Gemery, Ö. Gustafsson, K. Jerram, C. Johansson, A. Khortov, R. Mohammad, I. Semiletov, Evidence for an ice shelf covering the central Arctic Ocean during the penultimate glaciation. *Nature Communications*, **7**(1), 1-10 (2016).

Johnson, P. E., L. A. Young, S. Protopapa, B. Schmitt, L. R. Gabasova, B. L. Lewis, J. A. Stansberry, K. E. Mandt, O. L. White, Modeling Pluto's minimum pressure: implications for haze production. *Icarus*, **356**, 114070 (2021).

Kenyon, S. J., B. C. Bromley. Craters on Charon: Impactors from a collisional cascade among Trans-Neptunian Objects. *Planetary Science Journal*, **1**(2)**,** 40 (2020).

King, E. C., R. C. Hindmarsh, C. R. Stokes, Formation of mega-scale glacial lineations observed beneath a West Antarctic ice stream. *Nature Geoscience*, **2**(8), 585-588 (2009).

Kronberg, P., E. Hauber, M. Grott, S. C. Werner, T. Schäfer, K. Gwinner, B. Giese, P. Masson, G. Neukum, Acheron Fossae, Mars: Tectonic rifting, volcanism, and implications for lithospheric thickness. *Journal of Geophysical Research: Planets*, **112**(E4). E04005 (2007).

Lavvas, P., E. Lellouch, D. F. Strobel, M. A. Gurwell, A. F. Cheng, L. A. Young, G. R. Gladstone, A major ice component in Pluto's haze. *Nature Astronomy*, **5**(3), 289-297 (2021).

Lopes, R. M. C., K. L. Mitchell, E. R. Stofan, J. I. Lunine, R. Lorenz, F. Paganelli, R. L. Kirk, C. A. Wood, S. D. Wall, L. E. Robshaw, A. D. Fortes, C. D. Neish, J. Radebaugh, E. Reffet, S. J.

Ostro, C. Elachi, M. D. Allison, Y. Anderson, R. Boehmer, G. Boubin, P. Callahan, P. Encrenaz, E. Flamini, G. Francescetti, Y. Gim, G. Hamilton, S. Hensley, M. A. Janssen, W. T. K. Johnson, K. Kelleher, D. O. Muhleman, G. Ori, R. Orosei, G. Picardi, F. Posa, L. E. Roth, R. Seu, S. Shaffer, L. A. Soderblom, B. Stiles, S. Vetrella, R. D. West, L. Wye, H. A. Zebker, Cryovolcanic features on Titan's surface as revealed by the Cassini Titan Radar Mapper. *Icarus*, **186**(2), 395-412 (2007).

Lorenz, R. D., R. M. Lopes, F. Paganelli, J. I. Lunine, R. L. Kirk, K. L. Mitchell, L. A. Soderblom, E. R. Stofan, G. Ori, M. Myers, H. Miyamoto, J. Radebaugh, B. Stiles, S. D. Wall, C. A. Wood, the Cassini RADAR Team, Fluvial channels on Titan: initial Cassini RADAR observations. *Planetary and Space Science*, **56**(8), 1132-1144 (2008).

Lucchitta, B. K., Landslides in Valles Marineris, Mars. *Journal of Geophysical Research: Solid Earth*, **84**(B14), 8097-8113 (1979).

Malamud, U., H. B. Perets, G. Schubert, The contraction/expansion history of Charon with implications for its planetary-scale tectonic belt. *Monthly Notices of the Royal Astronomical Society*, **468**(1), 1056-1069 (2017).

Martin, C. R., Binzel, R. P., Ammonia-water freezing as a mechanism for recent cryovolcanism on Pluto. *Icarus*, **356**, 113763(2021).

Matsuyama, I., F. Nimmo, "Pluto's Tectonic Pattern Predictions" in *44th Annual Lunar and Planetary Science Conference*, No. 1719, p. 1399 (2013).

McKinnon, W.B., On the origin of the Pluto-Charon binary. *The Astrophysical Journal*, **344**, L41-L44 (1989).

McKinnon, W. B., C. R. Glein, T. Bertrand, A. R. Rhoden, "Formation, composition, and history of the Pluto system: A post-New Horizons synthesis" in *The Pluto System After New Horizons* (Univ. of Arizona, Tucson, 2021), pp. 507–543.

Mège, D., P. Masson, Amounts of crustal stretching in Valles Marineris, Mars. *Planetary and Space Science*, **44**(8), 749-781 (1996).

Moore, J. M., W. B. McKinnon, J. R. Spencer, A. D. Howard, P. M. Schenk, R. A. Beyer, F. Nimmo, K. N. Singer, O. M. Umurhan, O. L. White, S. A. Stern, K. Ennico, C. B. Olkin, H. A. Weaver, L. A. Young, R. P. Binzel, M. W. Buie, B. J. Buratti, A. F. Cheng, D. P. Cruikshank, W. M. Grundy, I. R. Linscott, H. J. Reitsema, D. C. Reuter, M. R. Showalter, V. J. Bray, C. L. Chavez, C. J. A. Howett, T. R. Lauer, C. M. Lisse, A. H. Parker, S. B. Porter, S. J. Robbins, K. Runyon, T. Stryk, H. B. Throop, C. C. C. Tsang, A. J. Verbiscer, A. M. Zangari, A. L. Chaikin, D. E. Wilhelms,. New Horizons Science Team, The geology of Pluto and Charon through the eyes of New Horizons. *Science*, **351**(6279), 1284-1293 (2016).

Morbidelli, A., D.C. Nesvorny, W. F. Bottke, S. Marchi, A re-assessment of the Kuiper belt size distribution for sub-kilometer objects, revealing collisional equilibrium at small sizes. *Icarus* **356**, 114256 (2021).

Morse, Z. R., G. R. Osinski, L.L. Tornabene, New morphological mapping and interpretation of ejecta deposits from Orientale Basin on the Moon. *Icarus* **299**, 253-271 (2018).

Nahm, A. L., R. A. Schultz, Evaluation of the orogenic belt hypothesis for the formation of the Thaumasia Highlands, Mars. *Journal of Geophysical Research: Planets*, **115**(E4) https://doi.org/10.1029/2009JE003327 (2010).

Nesvorný, D., Dynamical evolution of the early Solar System. *Annual Review of Astronomy and Astrophysics* **56**, 137-174 (2018).

Phillips, R. J., M. T. Zuber, S. E. Smrekar, M. T. Mellon, J. W. Head, K. L. Tanaka, N. E. Putzig, S. M. Milkovich, B. A. Campbell, J. J. Plaut, A. Safaeinili, R. Seu, D. Biccari, L. M. Carter, G. Picardi, R. Orosei, P. S. Mohit, E. Heggy, R. W. Zurek, A. F. Egan, E. Giacomoni, F. Russo, M. Cutigni, E. Pettinelli, J. W. Holt, C. J. Leuschen, L. Marinangeli, Mars north polar deposits: Stratigraphy, age, and geodynamical response. *Science*, **320**(5880), 1182-1185 (2008).

Protopapa, S., J. C. Cook, W. M. Grundy, D. P. Cruikshank, C. M. Dalle Ore, R. A. Beyer, "Surface composition of Charon" in *The Pluto System After New Horizons* (Univ. of Arizona, Tucson, 2021), pp. 433–456.

Reuter, D. C., S. A. Stern, J. Scherrer, D. E. Jennings, J. W. Baer, J. Hanley, Lisa Hardaway, A. Lunsford, S. McMuldroch, J. Moore, C. Olkin, R. Parizek, H. Reitsma, D. Sabatke, J. Spencer, J. Stone, H. Throop, J. V. Cleve, G. E. Weigle, L. A. Young, Ralph: A visible/infrared imager for the New Horizons Pluto/Kuiper Belt Mission. *Space Science Reviews*, **140**(1–4), 129–154 (2008).

Robbins, S. J., & Singer, K. N. (2021). Pluto and Charon Impact Crater Populations: Reconciling Different Results. *The Planetary Science Journal*, **2**(5), 192.

Rhoden, A. R., H. L. Skjetne, W. G. Henning, T. A. Hurford, K. J. Walsh, S. A. Stern, C. B. Olkin, J. R. Spencer, H. A. Weaver, L. A. Young, K. Ennico, and the New Horizons Team, Charon: A brief history of tides. *Journal of Geophysical Research: Planets*, **125**(7), e2020JE006449 (2020).

Robbins, S. J., K. N.Singer, V. J. Bray, P. Schenk, T. R. Lauer, H. A. Weaver, K. Runyon, W. B. McKinnon, R. A. Beyer, S. Porter, O. L. White, J. D. Hofgartner, A. M. Zangari, J. M. Moore, L. A. Young, J. R. Spencer, R. P. Binzel, M. W. Buie, B. J. Buratti, A. F. Cheng, W. M. Grundy, I. R. Linscott, H. J. Reitsema, D. C. Reuter, M. R. Showalter, G. L. Tyler, C. B. Olkin,

K. S. Ennico, S. A. Stern, the New Horizons LORRI, MVIC Instrument Teams, Craters of the Pluto-Charon system. *Icarus* **287**, 187-206 (2017).

Robbins, S. J., K. Runyon, K. N. Singer, V. J. Bray, R. A. Beyer, P. Schenk, W. B. McKinnon, W. M. Grundy, F. Nimmo, J. M. Moore, J. R. Spencer, O. L. White, R. P. Binzel, M. W. Buie, B. J. Buratti, A. F. Cheng, I. R. Linscott, H. J. Reitsema, D. C. Reuter, M. R. Showalter, G. L. Tyler, L. A. Young, C. B. Olkin, K. S. Ennico, H. A. Weaver, S. A. Stern Investigation of Charon's craters with abrupt terminus ejecta, comparisons with other icy bodies, and formation implications. *Journal of Geophysical Research: Planets*, **123**(1), 20-36 (2018).

Robbins, S. J., R. A. Beyer, J. R. Spencer, W. M. Grundy, O. L. White, K. N. Singer, J. M. Moore, C. M. Dalle Ore, W. B. McKinnon, C. M. Lisse, K. Runyon, C. B. Beddingfield, P. Schenk, O. M. Umurhan, D. P. Cruikshank, T. R. Lauer, V. J. Bray, R. P. Binzel, M. W. Buie, B. J. Buratti, A. F. Cheng, I. R. Linscott, D. C. Reuter, M. R. Showalter, L. A. Young, C. B. Olkin, K. S. Ennico, H. A. Weaver, S. A. Stern, Geologic landforms and chronostratigraphic history of Charon as revealed by a hemispheric geologic map. *Journal of Geophysical Research: Planets*, **124**(1), 155-174 (2019).

Schenk, P. M., R. A. Beyer, W. B. McKinnon, J. M. Moore, J. R. Spencer, O. L. White, K. Singer, O. M. Umurhan, F. Nimmo, T. R. Lauer, W. M. Grundy, S. Robbins, S. A. Stern, H. A. Weaver, L. A. Young, K. S. Ennico, C. Olkin, the New Horizons Geology and Geophysics Investigation Team, Breaking up is hard to do: Global cartography and topography of Pluto's mid-sized icy Moon Charon from New Horizons. *Icarus*, **315**, 124-145 (2018a).

Schenk, P. M., R. A. Beyer, W. B. McKinnon, J. M. Moore, J. R. Spencer, O. L. White, K. Singer, F. Nimmo, C. Thomason, T. R. Lauer, S. Robbins, O. M. Umurhan, W. M. Grundy, S. A. Stern, H. A. Weaver, L. A. Young, K. E. Smith, C. Olkin, the New Horizons Geology and

Geophysics Investigation Team, Basins, fractures and volcanoes: Global cartography and topography of Pluto from New Horizons. *Icarus*, **314**, 400-433 (2018b).

Schlichting, H. E., Runaway growth during planet formation: Explaining the size distribution of large kuiper belt objects. *The Astrophysical Journal,* **728**(1), 68 (2011).

Schlichting, H.E., E. O. Ofek, E. P. Nelan, A. Gal-Yam, M. Wenz, P. Muirhead, N. Javanfar, M. Livio, Measuring the abundance of sub-kilometer-sized Kuiper belt objects using stellar occultations. *The Astrophysical Journal,* **761**(2), 150 (2012).

Schlichting, H.E., C. I. Fuentes, D. E. Trilling, Initial planetesimal sizes and the size distribution of small Kuiper belt objects. *The Astronomical Journal,* **146**(2), 36 (2013).

Schulson, E. M., Brittle failure of ice. *Engineering fracture mechanics*, **68**(17-18), 1839-1887 (2001).

Sharp, R. P., Malaspina Glacier, Alaska. *Geological Society of America Bulletin*, **69**(6), 617-646 (1958).

Singer, K. N., W. B. McKinnon, B. Gladman, S. Greenstreet, E. B. Bierhaus, S. A. Stern, A. H. Parker, S. J. Robbins, P. M. Schenk, W. M. Grundy, V. J. Bray, R. A. Beyer, R. P. Binzel, H. A. Weaver, L. A. Young, J. R. Spencer, J. J. Kavelaars, J. M. Moore, A. M. Zangari, C. B. Olkin, T. R. Lauer, C. M. Lisse, K. Ennico, New Horizons Geology, Geophysics and Imaging Science Theme Team, New Horizons Surface Composition Science Theme Team, New Horizons Ralph and LORRI Teams, Impact craters on Pluto and Charon indicate a deficit of small Kuiper belt objects. *Science,* **363**(6430), 955-959 (2019).

Singer, K. N., O. L. White, B. Schmitt, E. L. Rader, S. Protopapa, W. M. Grundy, D. P. Cruikshank, T. Bertrand, P. M. Schenk, W. B. McKinnon, S. A. Stern, R. D. Dhingra, K. D. Runyon, R. A. Beyer, V. J. Bray, C. D. Ore, J. R. Spencer, J. M. Moore, F. Nimmo, J. T. Kean, L. A. Young,

C .B. Olkin, T. R. Lauer, H. A. Weaver, K. Ennico-Smith Large-scale cryovolcanic resurfacing on Pluto. *Nature communications*, *13*(1), 1-9 (2022).

Spencer, J. R., A. C. Barr, L. W. Esposito, P. Helfenstein, A. P. Ingersoll, R. Jaumann, C. P. McKay, F. Nimmo, J. H. Waite, "Enceladus: An active cryovolcanic satellite" in *Saturn from Cassini-Huygens* (Springer, Dordrecht, 2009), pp. 683-724.

Spencer, J. R., S. A. Stern, J. M. Moore, H. A. Weaver, K. N. Singer, C. B Olkin, A. J. Verbiscer, W. B. McKinnon, J. Wm. Parker, R. A. Beyer, J. T. Keane, T. R. Lauer, S. B. Porter, O. L. White,B. J. Buratti, M. R. El-Maarry, C. M. Lisse, A. H. Parker, H. B. Throop, S. J. Robbins,O. M. Umurhan, R. P. Binzel, D. T. Britt, M. W. Buie, A. F. Cheng, D. P. Cruikshank, H. A. Elliott,G. R. Gladstone, W. M. Grundy, M. E. Hill, M. Horanyi, D. E. Jennings, J. J. Kavelaars, I. R. Linscott, D. J. McComas, R. L. McNutt Jr., S. Protopapa, D. C. Reuter, P. M. Schenk, M. R. Showalter, L. A. Young, A. M. Zangari, A. Y. Abedin, C. B. Beddingfield, S. D. Benecchi, E. Bernardoni, C. J. Bierson, D. Borncamp, V. J. Bray, A. L. Chaikin, R. D. Dhingra, C. Fuentes, T. Fuse, P.L. Gay, S. D. J. Gwyn, D. P. Hamilton, J. D. Hofgartner, M. J. Holman, A. D. Howard, C. J. A. Howett, H. Karoji, D. E. Kaufmann, M. Kinczyk, B. H. May, M. Mountain, M. Pätzold, J. M. Petit, M. R. Piquette, I. N. Reid, H. J. Reitsema, K. D. Runyon, S. S. Sheppard, J. A. Stansberry, T. Stryk, P. Tanga, D. J. Tholen, D. E. Trilling, L. H. Wasserman, The geology and geophysics of Kuiper Belt object (486958) Arrokoth. *Science* **367**(6481), eaay3999 (2020).

Spencer, J. R., R. A. Beyer, S. J. Robbins, K. N. Singer, F. Nimmo, "The geology and geophysics of Charon" in *The Pluto System After New Horizons* (Univ. of Arizona, Tucson, 2021), pp. 395–412.

Stern, S. A., Collisional time scales in the Kuiper disk and their implications. *The Astronomical Journal*, **110**, 856 (1995).

Stern, S. A., L. M. Trafton. "On the atmospheres of objects in the Kuiper belt" in *The Solar System beyond Neptune* (Univ. of Arizona, Tucson, 2008), pp. 365-380.

Stern, S. A., R. Gladstone, A. Zangari, T. Fleming, D. Goldstein, Transient atmospheres on Charon and water–ice covered KBOs resulting from comet impacts. *Icarus*, **246**, 298-302 (2015).

Stofan, E. R., C. Elachi, J. I. Lunine, R. D. Lorenz, B. Stiles, K. L. Mitchell, S. Ostro, L. Soderblom, C. Wood, H. Zebker, S. Wall, M. Janssen, R. Kirk, R. Lopes, F. Paganelli, J. Radebaugh, L. Wye, Y. Anderson, M. Allison, R. Boehmer, P. Callahan, P. Encrenaz, E. Flamini, G. Francescetti, Y. Gim, G. Hamilton, S. Hensley, W. T. K. Johnson, K. Kelleher, D. Muhleman, P. Paillou, G. Picardi, F. Posa, L. Roth, R. Seu, S. Shaffer, S. Vetrella, R. West, The lakes of Titan. *Nature*, **445**(7123), 61-64 (2007).

Trafton, L., S. A. Stern, G. R. Gladstone, The Pluto-Charon system: The escape of Charon's primordial atmosphere. *Icarus*, **74**(1), 108-120 (1988).

Waller, R. I., The influence of basal processes on the dynamic behaviour of cold-based glaciers. *Quaternary International*, **86,** 117–128 (2001).

Watkins, J. A., Ehlmann, B. L., & Yin, A. (2015). Long-runout landslides and the long-lasting effects of early water activity on Mars. *Geology*, **43**(2), 107-110.

Watkins, J. A., Ehlmann, B. L., & Yin, A. (2020). Spatiotemporal evolution, mineralogical composition, and transport mechanisms of long-runout landslides in Valles Marineris, Mars. *Icarus*, **350**, 113836.

Watters, T. R., M. M. Selvans, M. E. Banks, S. A. Hauck, K. J. Becker, M. S. Robinson, Distribution of large-scale contractional tectonic landforms on Mercury: Implications for the origin of global stresses. *Geophysical Research Letters*, **42**(10), 3755-3763 (2015).

Weertman, J., Creep deformation of ice. *Annual Review of Earth and Planetary Sciences*, **11**(1), 215-240 (1983).

Weiss, D. K., and J. W. Head, Formation of double-layered ejecta craters on Mars: A glacial substrate model. *Geophysical Research Letters*, **40**(15), 3819–3824 (2013).

Yamashita, Y., M. Kato, M. Arakawa, Experimental study on the rheological properties of polycrystalline solid nitrogen and methane: Implications for tectonic processes on Triton. *Icarus*, **207**(2), 972-977 (2010).

Yin, A. Mode of Cenozoic east-west extension in Tibet suggesting a common origin of rifts in Asia during the Indo-Asian collision. *Journal of Geophysical Research: Solid Earth*, **105**(B9), 21745-21759 (2000).

Yin, A. Structural analysis of the Valles Marineris fault zone: Possible evidence for large-scale strike-slip faulting on Mars. *Lithosphere*, ***4*(4)**, 286-330 (2012a).

Yin, A. An episodic slab-rollback model for the origin of the Tharsis rise on Mars: Implications for initiation of local plate subduction and final unification of a kinematically linked global plate-tectonic network on Earth. *Lithosphere*, ***4*(6)**, 553-593 (2012b).

Yin, A., R. T. Pappalardo, Gravitational spreading, bookshelf faulting, and tectonic evolution of the South Polar Terrain of Saturn's moon Enceladus. *Icarus*, **260**, 409-439 (2015).

Yin, A., S. Moon, M. Day, Landform evolution of Oudemans crater and its bounding plateau plains on Mars: Geomorphological constraints on the Tharsis ice-cap hypothesis. *Icarus*, **360**, 114332 (2021).

Youdin, A. N., J. Goodman, Streaming instabilities in protoplanetary disks. *Astrophysical Journal*, **620**(1), 459 (2005).

Zhang, X., D. F. Strobel, H. Imanaka, Haze heats Pluto's atmosphere yet explains its cold temperature. *Nature*, **551**(7680), 352-355 (2017)

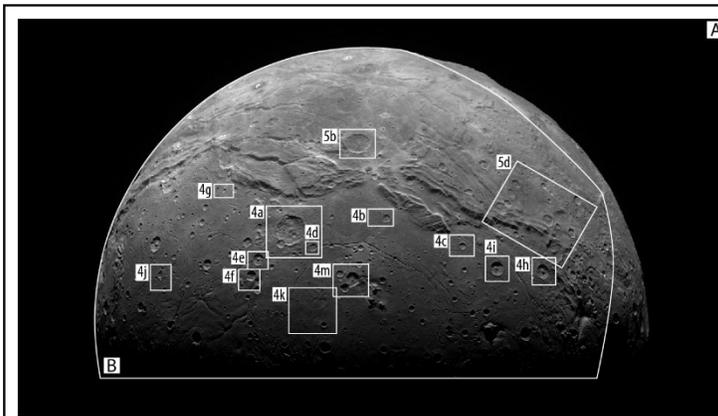
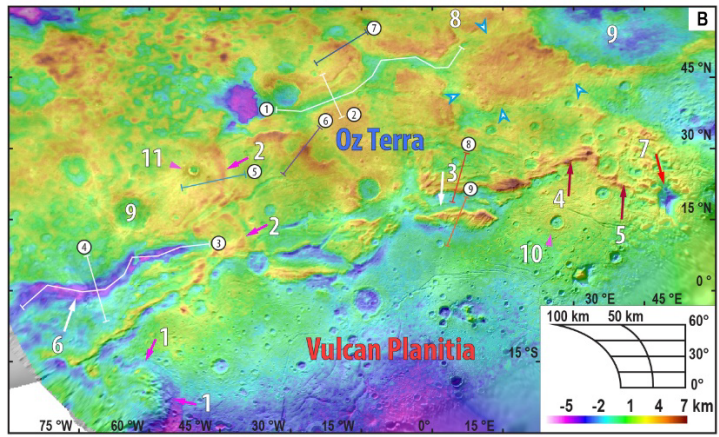
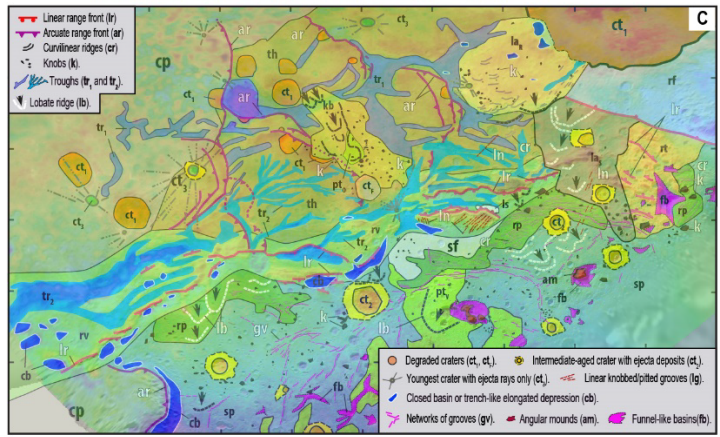
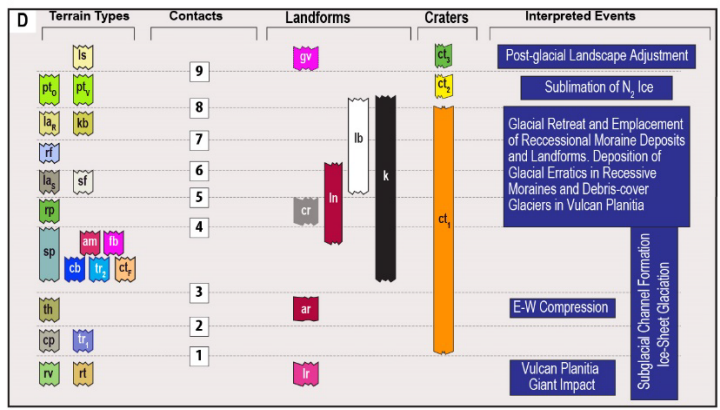

**Fig. 1. Geomorphologic map and chronostratigraphy of Charon's encounter hemisphere**. (A) Enhanced color view of Charon that combines blue, red and infrared images taken by the New Horizons spacecraft's Ralph/Multispectral Visual Imaging Camera (MVIC) (Moore et al., 2016). (B) A digital elevation model of Charon's encounter hemisphere at a horizontal resolution of 300 m/pixel overlying the Charon New Horizons LORRI MVIC Global Mosaic 300 m v1 (Schenk et al., 2018a). (C) Geomorphological map of Charon's encounter hemisphere created in this study. Description of terrain units is listed on the column to the right. (D) The interpreted chronological sequence of the mapped landform units in the left and possible geological processes responsible for their formation on the right. Numbers in (D) are the contacts between landform units. See text for detailed description of landform units.

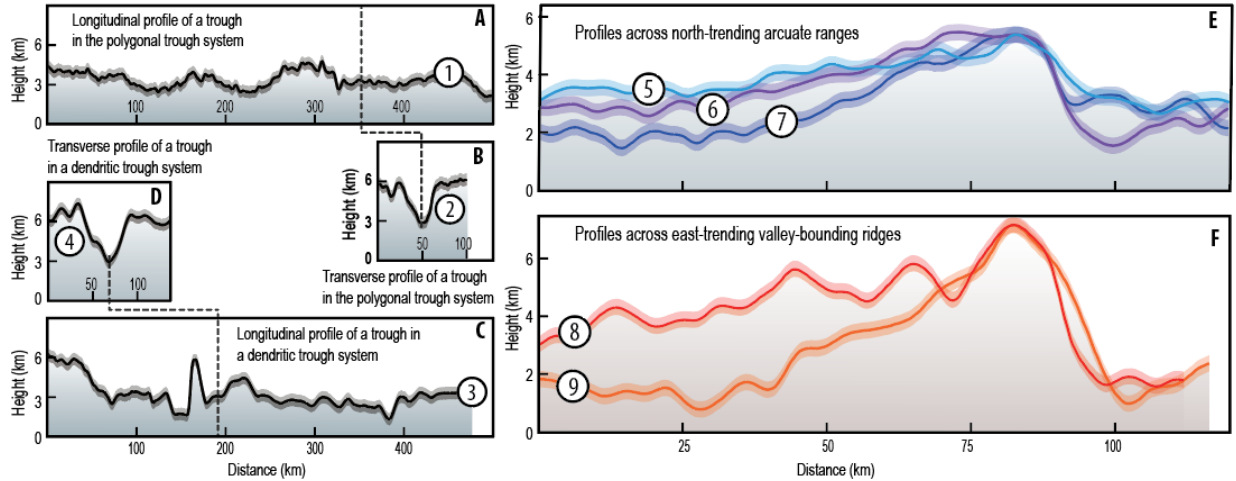

**Fig. 2. Topographic profiles of ranges, ridges, and troughs**. (A) Longitudinal and (B) transverse profiles of a linear trough in the polygonal trough network (unit *tr₁*). (C) Longitudinal and (D) transverse profiles of a curvilinear trough in a dendritic trough network (unit *tr₂*). (E) Topographic profiles across north-trending arcuate ranges. (F) Topographic profiles across ridges in the eastern ridge-bounding valley zone (unit *rv*). The widths of the topographic profile lines represent the uncertainties quoted from Schenk et al. (2018a). Digital elevation model data are from Schenk et al. (2018a).

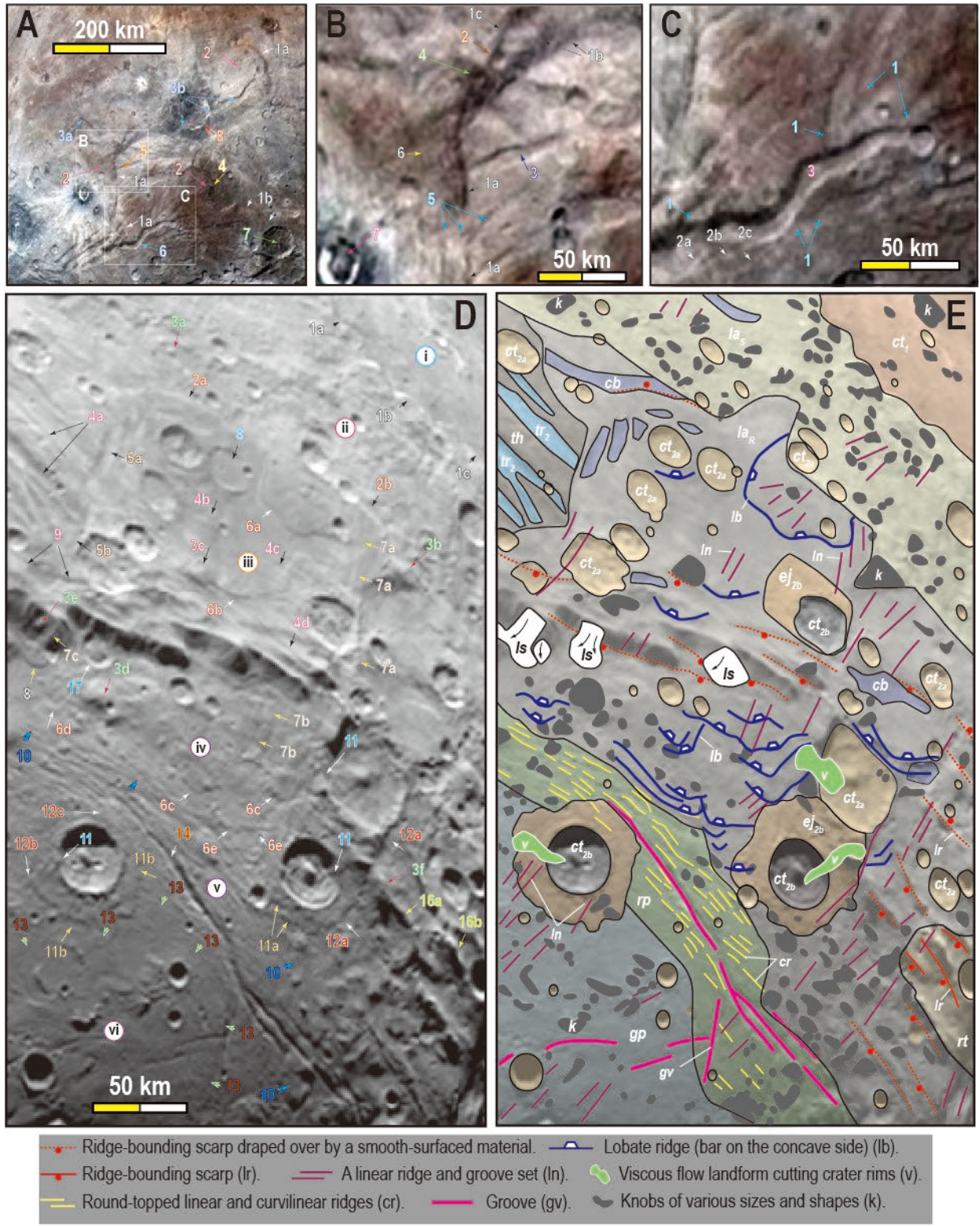

**Fig. 3. Morphological associations of the mapped landform units**. (A) Key morphological features and their spatiotemporal relationships in the troughed highland terrain (*th*). East-facing

scarps (feature 1a) bounding north-trending ranges that have longer west-sloping flanks (feature 2) and shorter eastern range flanks. The ranges are separated by parallel north-trending troughs (feature 3a). An arcuate range front (1b) truncates both crater rims (feature 4) and an older set of east-trending troughs (feature 5). The north-trending ranges themselves are crosscut by a younger dendritic network of troughs (feature 6). The ranges are also truncated (see feature 1b) by a younger crater (feature 7) filled by a ridged-plain material. A degraded crater with a dark-material apron (feature 8) is crosscut by an east-trending trough (feature 3b). (B) A zoom-in view of an east-facing scarp zone that consists of a single scarp trace in the south (feature 1a) and multiple scarp traces (features 1b and 1c) in the north. The scarp zone truncates a crater (features 2) and troughs (feature 3), but the zone itself is crosscut by younger craters (e.g. features 4) and superposed over by a younger dendritic trough system (feature 5). The scarp-bounded range crest displays a set of range-front-parallel ridges (feature 6). The scarp-bounded west-sloping range flank is draped over by a younger crater surrounded by proximal darker-toned and distal lighter-toned ejecta deposits (feature 7). (C) A zoom-in view of a dendritic network of troughs ($tr_2$) that exhibits hanging-valley morphologies (feature 1) and terrace-rise-like scarps (features 2a-2c) along the margin of a meandered, steep-walled trough (feature 3). (D) Key landforms across the topographic boundary zone between Oz Terra and Vulcan Planitia. Dorothy crater (feature *i*) is characterized by a smooth-surfaced basin floor and lies next to the rough-surfaced lobate-apron terrain ($la_R$) on Oz Terra (feature *ii*). The smooth-surface lobate-apron terrain (feature *iii*) extends from Oz Terra to Vulcan Planitia (feature *iv*). Features 1a-1c mark the contact between Dorothy crater and unit $la_R$, whereas features 2a and 2b marks the contact between unit $la_R$ and the smooth-surfaced lobate-apron terrain (unit $la_S$ and feature *iii*). Features 3a-3b are examples of mapped knobs in unit $la_R$, which are up to >20 km in the longest dimension. Features 3c and 3d are examples of knobs mapped in unit $la_S$, which have smoother surfaces than those in unit $la_R$ (cf. feature 3b). Note that a knob is draped over a south-facing scarp (feature 3e). Feature 4a marks a zone of northwest-trending linear troughs mapped as unit $tr_2$ that are crosscut by a northeast-trending linear depression (feature 5a) and a similarly trending scarp (feature 5b). The troughs ($tr_2$) and their bounding walls are draped over by unit $la_S$ (features 4b-4d). Unit $la_S$ hosts south-convex lobate aprons (features 6a, 6b and 6c), south-convex lobate ridges (features 6d and 6e), north-trending linear ridges and grooves (features 7a and 7b), and craters filled by flat smooth-surfaced materials (feature 8; also see description of this same feature in Robbins et al., 2019). A north-trending ridge (feature 7c on the left-central edge of the image) exposed on the surface of a south-facing scarp terminates downward at a lobate apron (feature 8). The lack of a corresponding breakaway topographic feature rules out the apron landform to be a landslide. Feature 9 shows a possible degraded crater cut by a south-facing scarp, which likely represents the trace of a normal fault. Feature 10 represents the contact between the smooth-surfaced lobate-apron unit (features *iii* and *iv*) and the ridged plain terrain (feature *v*). Feature 11 are flow-like features that cut across crater rims and extend into the crater-basin floors. Layered impact deposits for one crater are deformed by northwest-trending ridges (feature 11a), whereas layered impact deposits for another crater (feature 11b) are truncated (feature 12b) and overprinted (feature 12c) by a zone of parallel ridges (*cr*) in the ridged plains terrain (*rp*). The boundary between ridged-plain terrain (*rp* and feature *v*) and the smooth-surfaced lobate-apron terrain (unit $la_S$ and feature *vi*) is transitional (see feature 13). This contact is crosscut by younger grooves (feature 14). In the southwestern corner of the image, northwest-trending smooth-surfaced linear ridges (feature 16a) are locally superposed by knobs (feature 3f). In contrast, ridges with the same trend directly to the southeast display rougher surfaces (feature 16b). A partially filled crater lies at the base of a range bounding scarp (feature 17). (E) Geomorphological map that summarizes all the aforementioned landforms shown in (D). Abbreviated map symbols are: *k*, knobs; *ln*, sets of linear ridges and grooves; *v*, viscous flow features cutting across crater-rim ridges and extending into crater basin floors; *cr*, sets of parallel linear and curvilinear ridges; *gv*, grooves; and *lr*, linear ridge. All other landform units are defined in Fig. 1. Note that the geomorphological features superposed by the smooth-surfaced lobate-apron unit such as the east-trending scarps and northwest-trending ridges are shown by dashed lines.

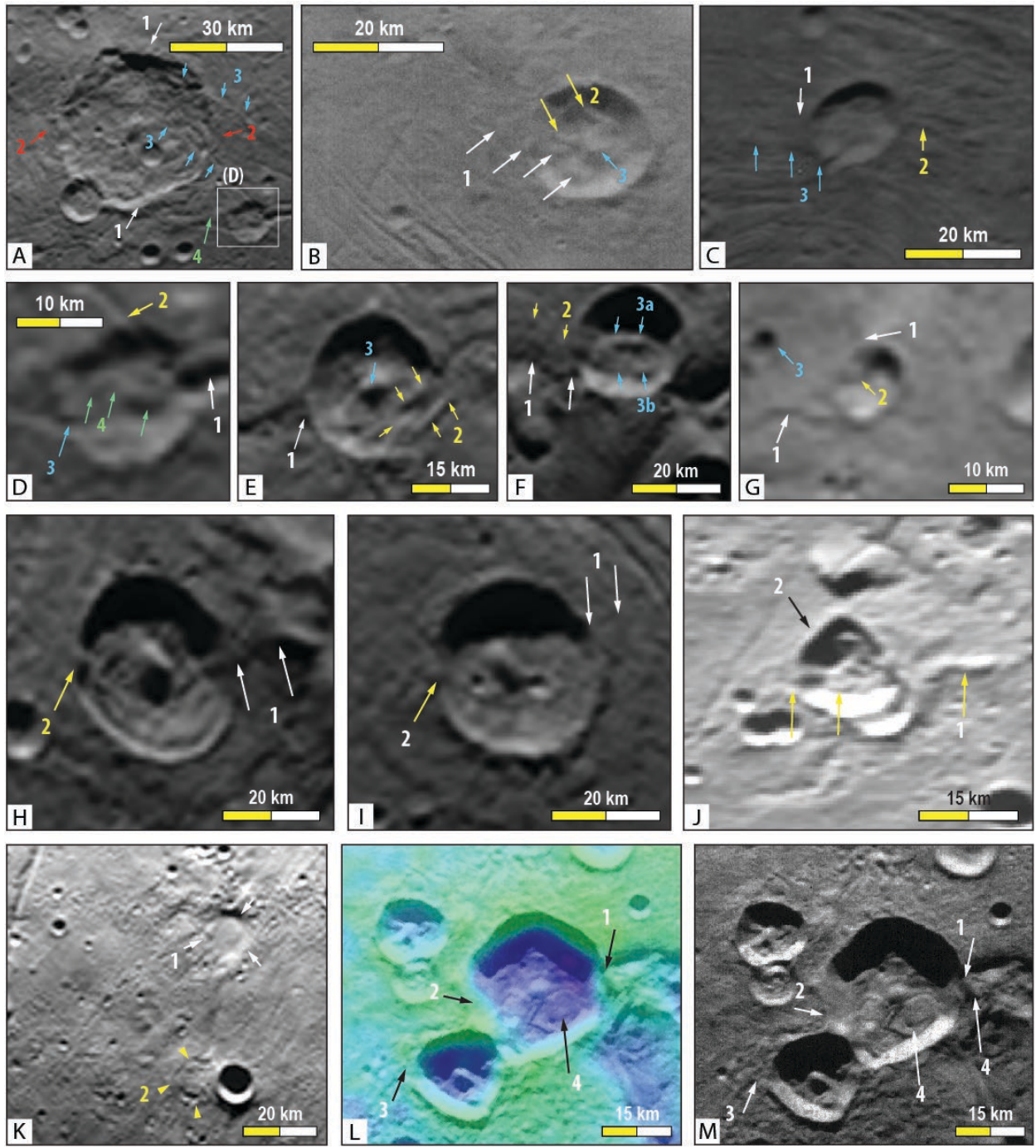

**Fig. 4. Modified craters in Vulcan Planitia.** (A) A crater with a partially preserved rim ridge marked as feature 1 along its northern and southern edges. In contrast, its western and eastern rims are bounded by flat plains where the rim ridge is missing (feature 2). The impact basin is cut across by a series of northwest trending ridges and grooves (feature 3). The crater floor and crater-

bounding plains share the same hummocky surface texture, which requires a post-cratering resurfacing process to create the textured surfaced (image ID mp1_0299180334_0x530_sci at a resolution of 622 m/pixel). (B) A crater with its western rim ridge breached and draped over by an 8-10 km wide gentle-sloped, smooth-surfaced material (feature 1). Parallel east-trending ridges (feature 2) and central mound (feature 3) are present on the crater floor (image ID lor_0299180418_0x630_sci_3 at a resolution of 154 m/pixel). (C) A crater with its western and eastern rims superposed by 8-10 km wide grooved materials (features 1 and 2). A wider ridge characterized by the occurrence of higher mounds at two ends cut across the crater rim (feature 3) (lor_0299175682_0x630_sci_4 at a resolution of 410 m/pixel). (D) A zoom-in view of the smaller crater shown in (A). The crater rim is breached in three places (features 1, 2, and 3) by 2-5 km wide grooves. The crater floor displays curvilinear ridges that terminate at the ends of the crater-rim-cutting grooves (image ID mp1_0299180334_0x530_sci at a resolution of 622 m/pixel). (E) A crater is breached by grooves and ridges across its western and eastern rims (feature 1 and 2). The crater floor displays a well-defined central peak (feature 3) (image ID mp1_0299180334_0x530_sci at a resolution of 622 m/pixel). (F) A crater with its western rim superposed by an 8-12-km wide, hummocky-surface material (feature 1). The flat top of the ridge is superposed by narrow ridges that are parallel to the general trend of the hosting ridge (feature 2). The hosting ridge splits into two braches as it enters the crater floor (feature 3a and 3b) (image ID mp1_0299180334_0x530_sci at a resolution of 622 m/pixel). (G) A crater that is partially filled by a sheet-like smooth-surfaced material with lobate-shaped scarps (feature 1 and 2). The sheet-like landform appears to have been sourced from a smaller, rampart style impact crater from the west (feature 3) (image ID mp1_0299180334_0x530_sci at a resolution of 622 m/pixel). (H) A complex crater with its eastern rim (feature 1) and western rim (feature 2) draped over by ridges

(image ID mp1_0299180334_0x530_sci at a resolution of 622 m/pixel). (I) A crater with its eastern rim (feature 1) and western rim (feature 2) breached and draped over by 7-km wide round-topped ridges (image ID mp1_0299180334_0x530_sci at a resolution of 622 m/pixel). (J) An east-trending ridge system is superposed over a quasi-triangular-shaped crater. Image is cropped out from the Charon New Horizons LORRI MVIC Global Mosaic 300 m v1 created by Schenk et al. (2018a). (K) Feature 1 shows a circular depression with a smooth floor that may represent a partially filled crater basin. Feature 2 shows a circular rim ridge and a circular groove, which could be the relics of a filled crater. (Image ID lor_0299180424_0x630_sci_3 at a resolution of 622 m/pixel). (L) An image cropped out from the global digital elevation model of Schenk et al. (2018a) at a resolution of 300 m/pixel and (M) an image of the same area cropped out from a larger image with the identification number of ID lor_0299180421_0x630_sci_3 at a resolution of 154 m/pixel. In both images, features 1, 2, and 3 indicate the locations where crater rims are breached by linear depressions, and feature 4 shows a tongue-shaped material that is superposed over the crater wall and the crater floor.

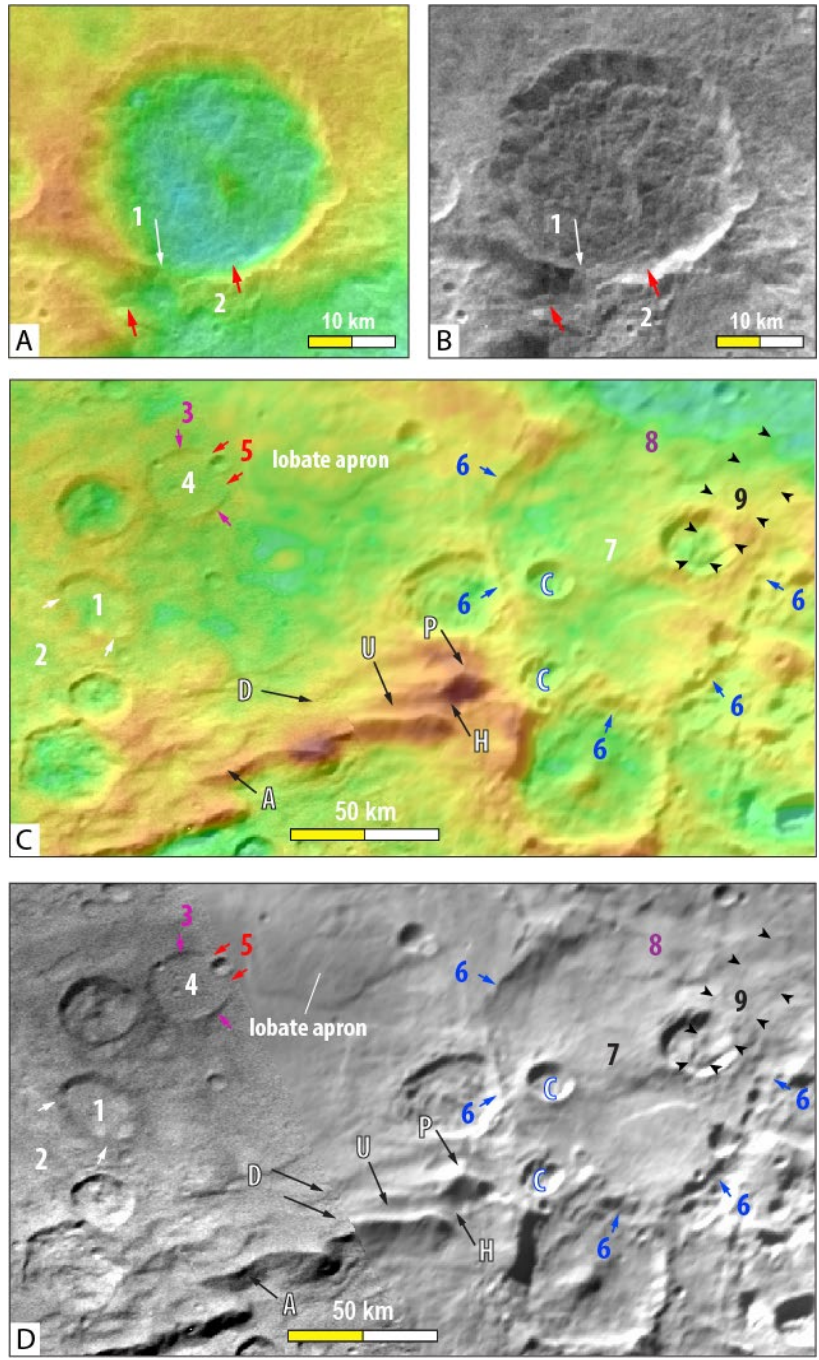

**Fig. 5. Modified craters in Oz Terra.** (A)-(B) An image cropped out from the global digital elevation model of Schenk et al. (2018a) at a resolution of 300 m/pixel (A) and an image of the same area cropped out from a larger image with the identification number of ID lor_0299180412_0x630_sci_3 at a resolution of 154 m/pixel (B). (C)-(D) An image cropped out from the global digital elevation model of Schenk et al. (2018a) at a resolution of 300 m/pixel (C)

and an image cropped out from the Charon New Horizons LORRI MVIC Global Mosaic 300 m v1 (Schenk et al., 2018a) (D).

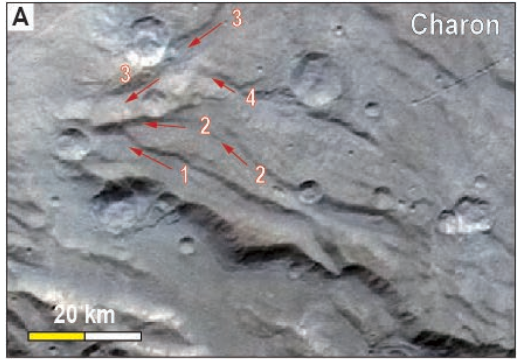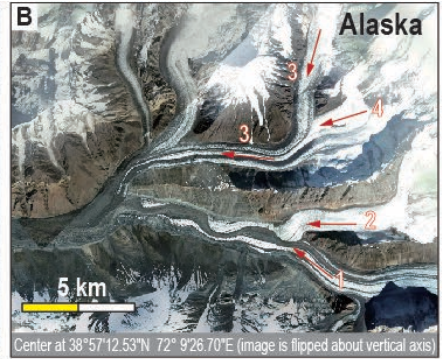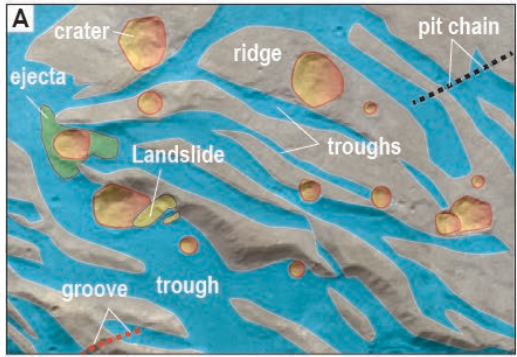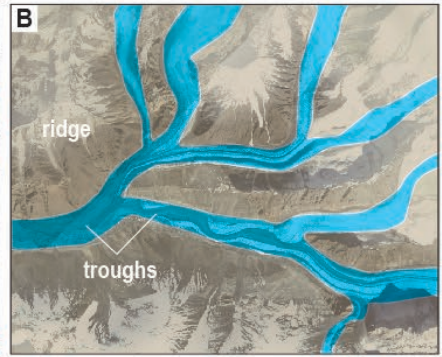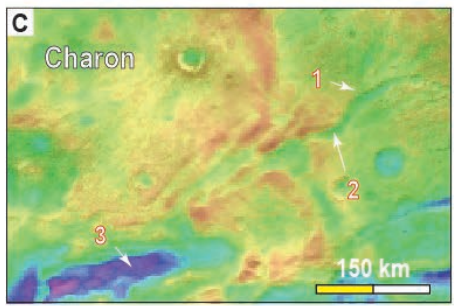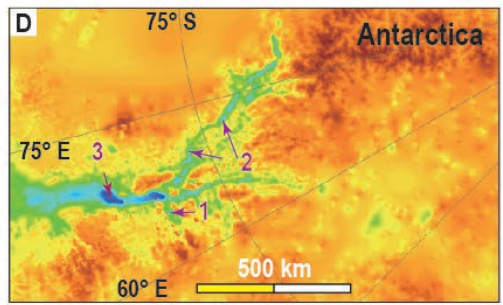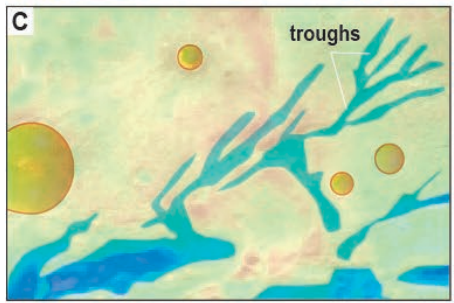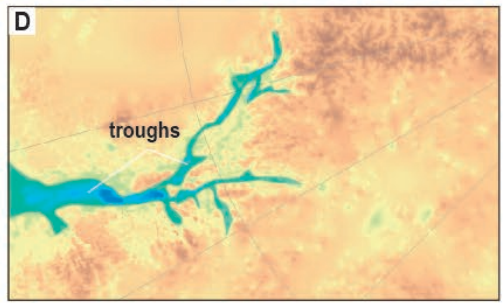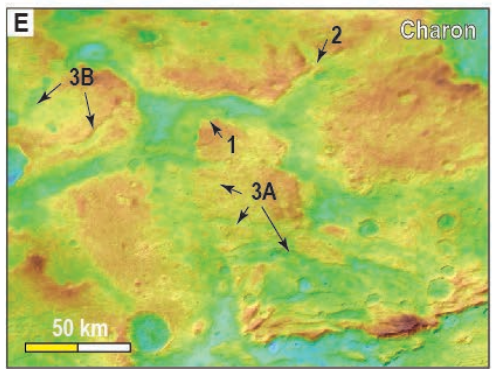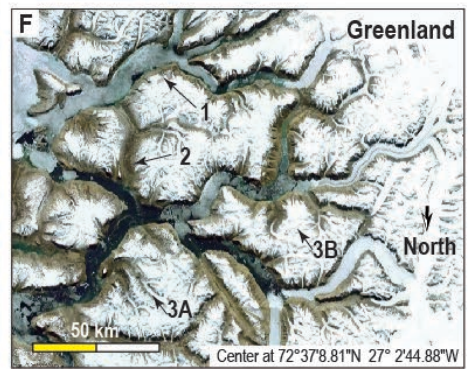

**Fig. 6. Comparison of Charon's dendritic and polygonal trough systems with possible analogues that have similar shapes and known origins from Earth**. (A)-(B) Comparison of dendritic trough systems on Charon (A) (image ID PIA 19967 in NASA's Photojournal Image Database) on Earth (B) from the Karakoram Mountains of central Asia taken from Google Earth$^{TM}$. Features 1 to 4 in both images are corresponding landforms with similar shapes. (C)-(D) Comparison of digital elevation models (DEMs) of a dendritic trough network on Charon (C) from Schenk et al. (2018a) and a subglacial dendritic network of troughs in Antarctica (D) with data from Fretwell et al. (2013). Note that both images display hanging-valleys (feature 1), undulating longitudinal profiles (feature 2), and overdeepenings along the main trough trunks (feature 3). (E)-(F) Comparison of polygonal networks of wide (10s km) and long (100s km) troughs on Charon (E) with the digital elevation model from Schenk et al. (2018a) and on Earth with the Google Earth$^{TM}$ image from a glaciated region of Greenland (F). Note the similarities in size and shape of the troughs (feature 1), the presence of hanging valleys terminating at main trunks (feature 2), and minor trough networks on the surface of trough-bounded plateaus (features 3A and 3B).

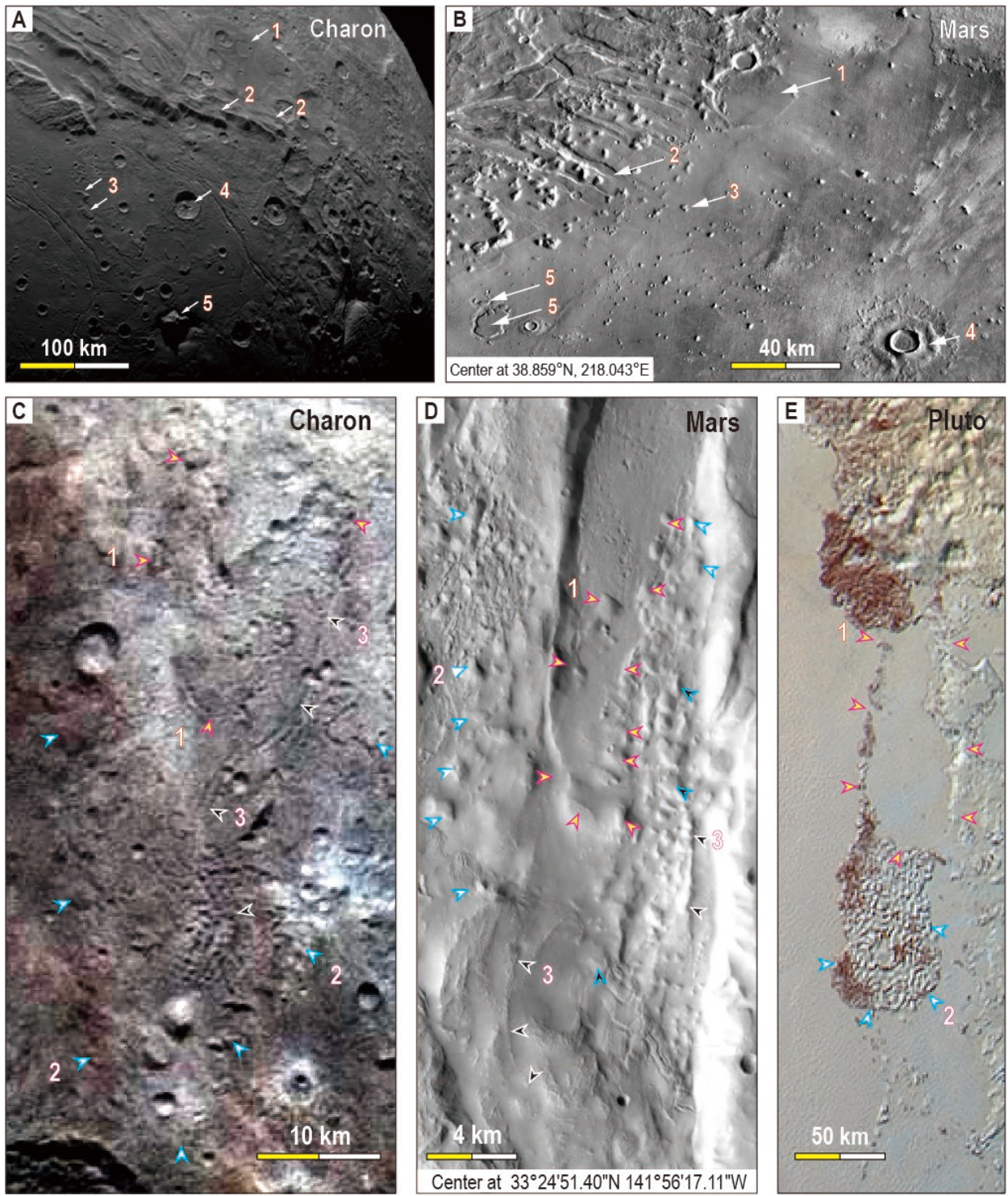

**Fig. 7. Comparison of Charon's landforms with possible analogues that have similar shapes and known origins from Mars and Pluto.** (A)-(B) show comparison of troughed highlands and their knobbed foreland plains on Charon (A) (image ID mp1_0299180334_0x530_sci at a

resolution of 622 m/pixel) and Mars (B) (THEMIS IR daytime image at a resolution of 100 m/pixel). Feature 1, a partially filled crater basin; feature 2, steep-walled U-shaped valleys cutting across the highlands; feature 3, scattered knobs in the foreland plains; feature 4, rampart crater; and feature 5, moated mounds. (C)-(E) are images from Charon (C) (image ID PIA 19967 in NASA's Photojournal Image Database), Mars (D) (THEMIS IR daytime image at a resolution of 100 m/pixel), and Pluto (E) (image cropped out from the Pluto Global Mosaic (Schenk et al., 2018b) that show similar lobate features and distributed knob trains that define the rim of the lobes.

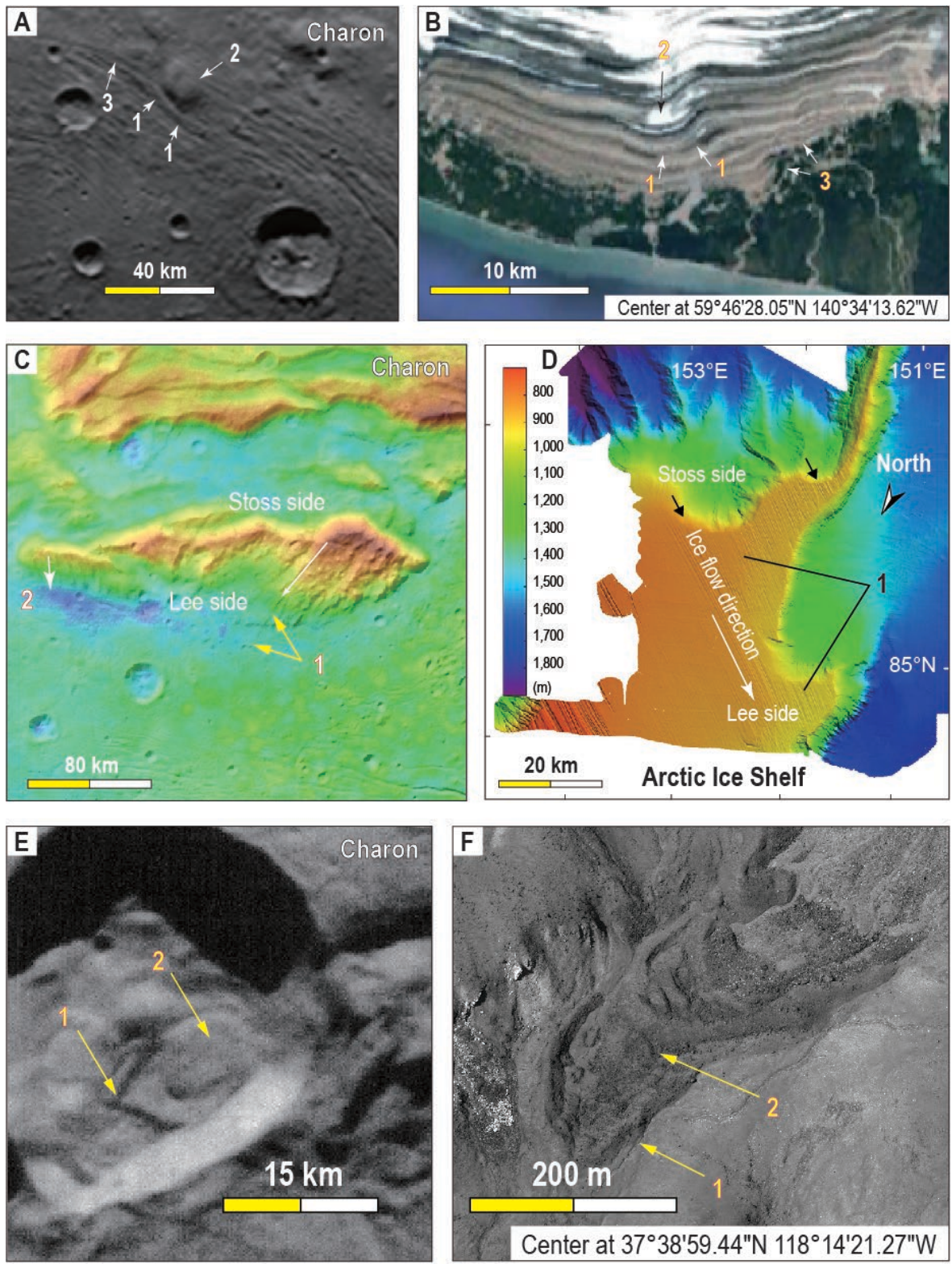

**Fig. 8. Comparison of Charon's landforms with possible analogues that have similar shapes and known origins from Earth.** (A)-(B) Comparison ridged plains on Charon (A) (image ID

mp1_0299180334_0x530_sci at a resolution of 622 m/pixel) and ridged plains in the foreland region of the Malaspina Glacier in Alaska on Earth (B) with the image cropped out from Google Earth[TM]. The parallel ridges in front of the Malaspina glacier were formed by folding of debris-bearing glaciers (Sharp, 1958). Note the similarities in the shape of the curved ridges (feature 1) that are deflected around local topographic heights (feature 2) and the lateral merging of the ridges (feature 3). (C)-(D) Comparison of digital elevation models of striated quasi-planar ridge flanks on Charon (C) (Schenk et al., 2018a) and on Earth (D) from an Arctic ice shelf adopted from Jakobsson et al. (2016). (E)-(F) Comparison of lobate ridges on Charon that filled a crater floor and lobate ridges composed of rock glaciers that filled a glaciated valley in the White Mountains of California (image cropped out from Google Earth[TM]).

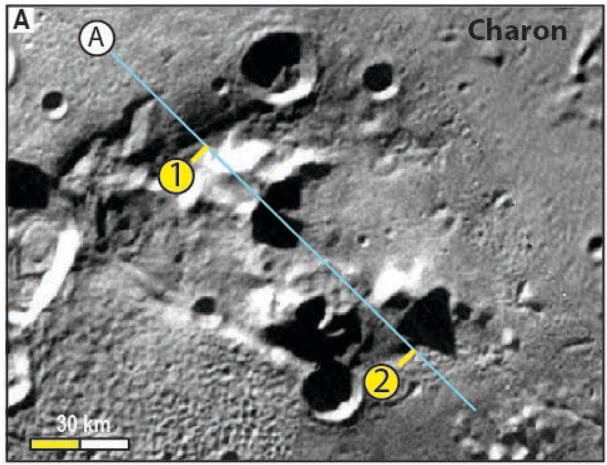
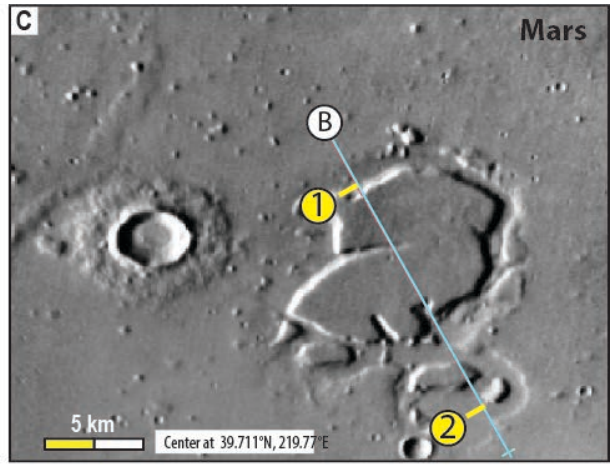
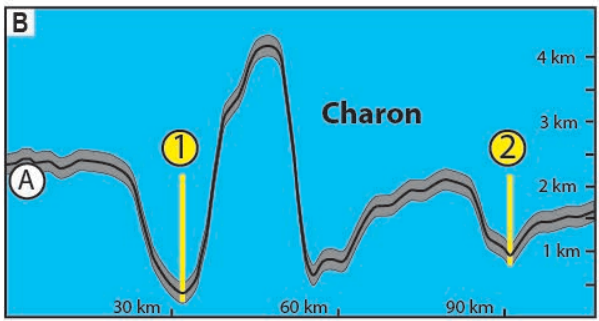
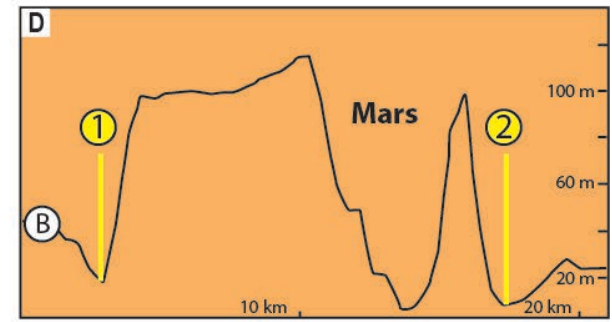
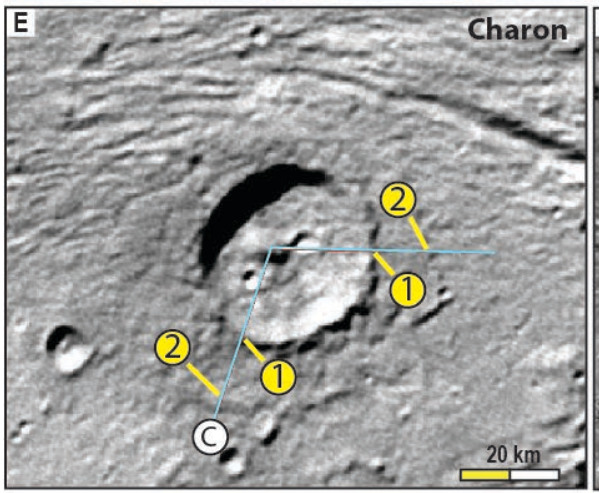
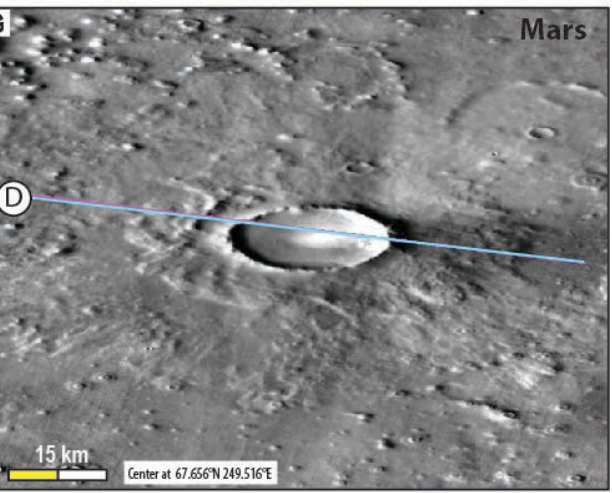
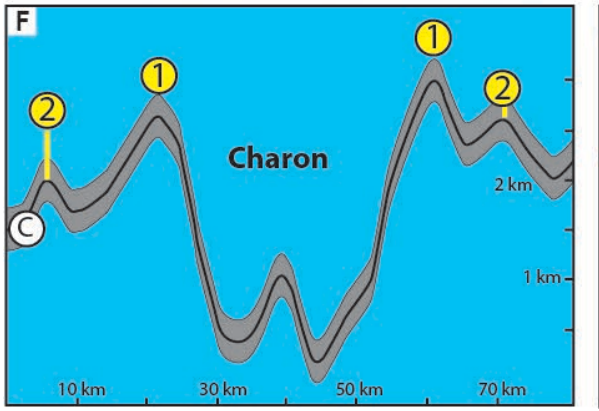
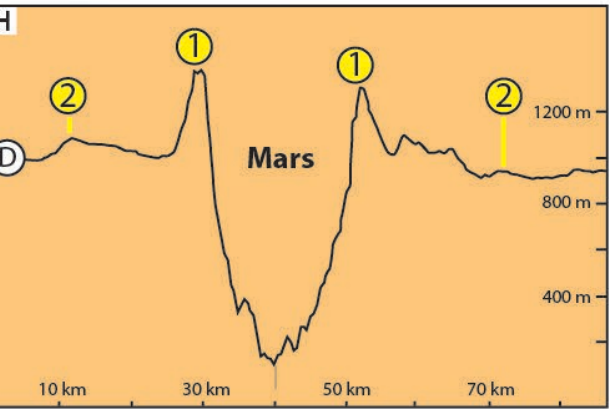

**Fig. 9. Comparison of moated mounds and rampart craters on Charon and Mars**. (A)-(B) An image of a moated mound complex on Charon cropped out from the Charon New Horizons LORRI MVIC Global Mosaic 300 m v1 (Schenk et al., 2018a) and a topographic profile obtained from the Charon digital elevation model created by Schenk et al. (2018a). Features 1 and 2 indicate the positions of the moat. (C)-(D) An image of moated mound complex on Mars (THEMIS IR daytime image at a resolution of 100 m/pixel) and a topographic profile constructed using the blended High Resolution Stereo Camera (HRSC) data and Mars Orbiter Laser Altimeter (MOLA) digital elevation model (DEM) at a resolution of 200 m/pixel (Fergason et al., 2018). Features 1 and 2 indicate the positions of the moat. (E)-(F) An image of a rampart crater on Charon cropped out from the Charon New Horizons LORRI MVIC Global Mosaic 300 m v1 (Schenk et al., 2018a) and a topographic profile obtained from the Charon digital elevation model created by Schenk et al. (2018a). Feature 1 indicates the rim ridge and feature 2 indicates the ejecta-terminus ridge. (G)-(H) An image of rampart crater on Mars (THEMIS IR daytime image at a resolution of 100 m/pixel) and a topographic profile constructed using the blended High Resolution Stereo Camera (HRSC) data and Mars Orbiter Laser Altimeter (MOLA) digital elevation model (DEM) at a resolution of 200 m/pixel (Fergason et al., 2018). Feature 1 indicates the rim ridge and feature 2 indicates the ejecta-terminus ridge.

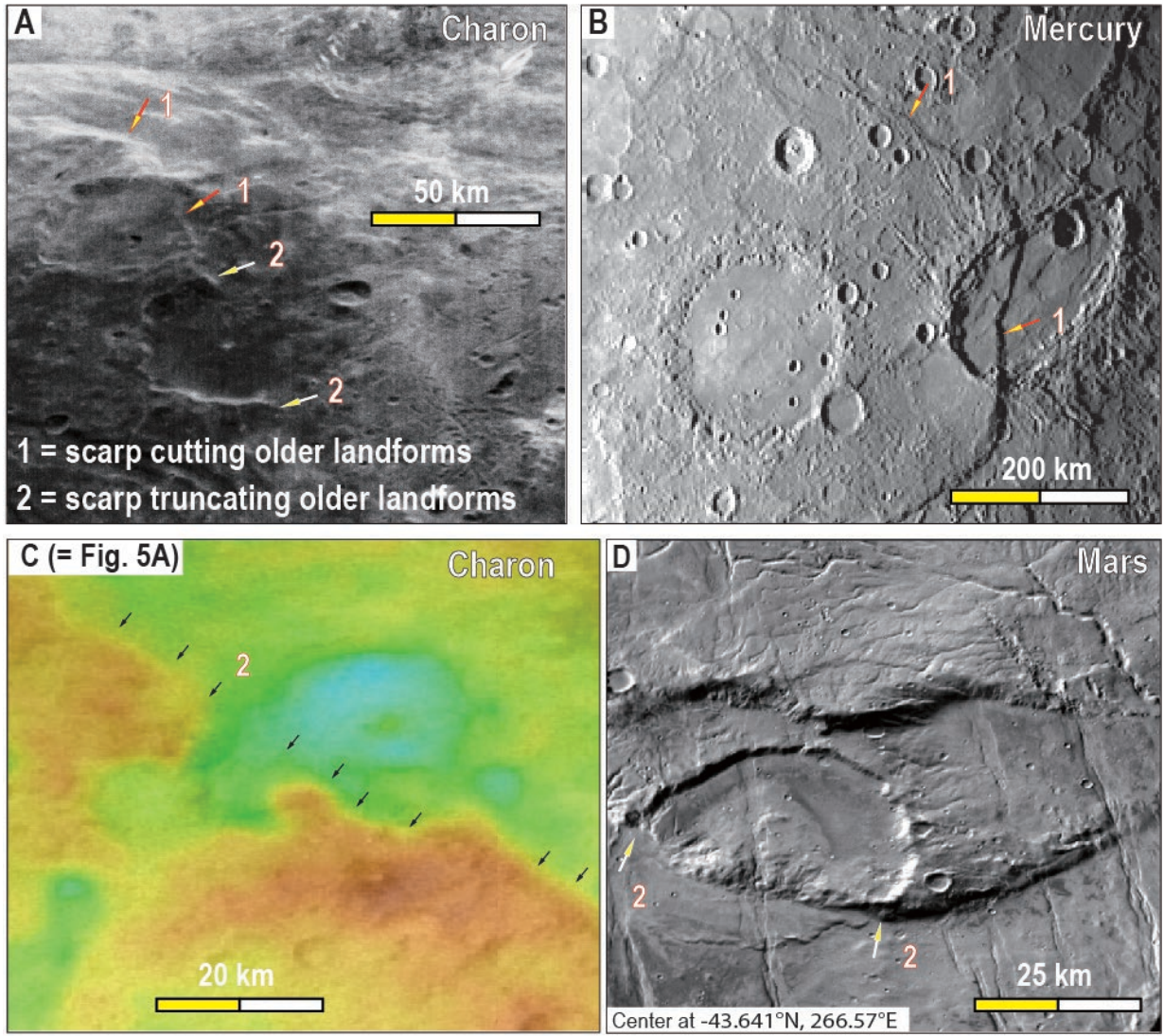

**Fig. 10. Comparison of deformation-modified craters on Charon, Mercury, and Mars.**

(A)-(D) Comparison of crater rim ridges either cut or truncated by scarps among those exposed on Charon, Mercury, and Mars. The Charon image in (A) is cropped out from image lor_0299180409_0x630_sci_3 at a resolution of 154 m/pixel, the Mercury image showing Beagle Rupes in (B) is from PIA10939 in the NASA's photojournal database, and the Mars image in (D) is cropped out from the CTX global mosaic displayed in JMARS.

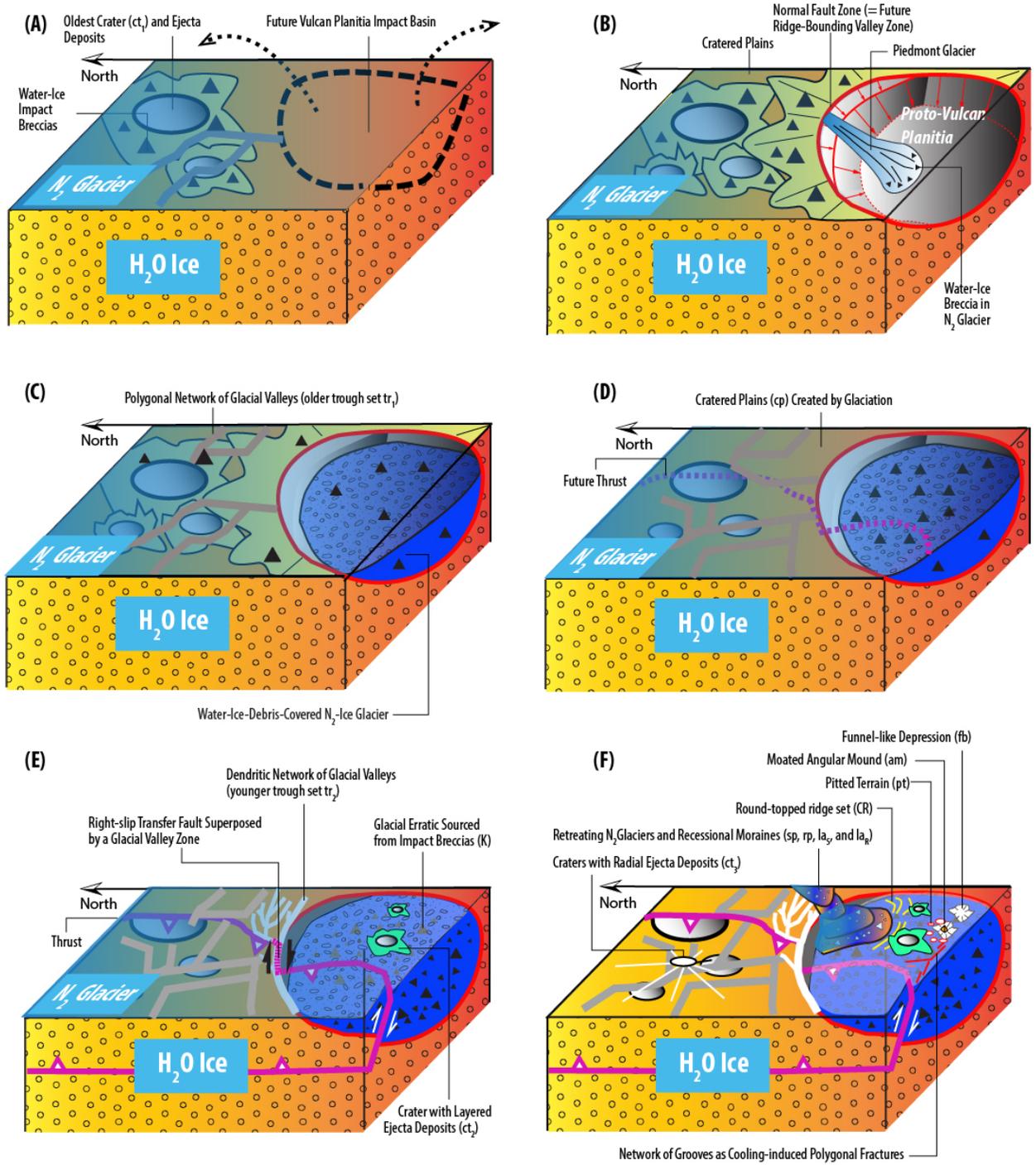

**Fig. 11. Landscape evolution of Charon's encounter hemisphere.** (A) Intensely bombarded surface of Charon immediately after its formation is characterized by larger craters (D>50 km) with water-ice breccia-bearing ejecta blankets. Although not shown, glaciation could have been coeval during the early intense bombardment. (B) The early intense bombardment by impacts immediately after the formation of Charon was responsible for the formation of proto-Vulcan Planitia, which was deeper than its present depth and surrounded by impact-induced normal faults and fault-bounded grabens. Again, glaciation could have occurred during the formation of proto-Vulcan Planitia. (C)-(D) Glaciation on the highlands of Oz Terra removed the older impact breccias, eroded the older crater-rim ridges, and infilled the older crater basins. Meanwhile, heating from below and basal pressure due to the weight of the $N_2$ ice sheet above caused the development of subglacial $N_2$–liquid channels. The removed impact breccias from Oz Terra were transported by $N_2$ glaciers to Vulcan Planitia. The progressive accumulation of water-ice blocks and their potentially smaller densities than the viscous $N_2$ ice created a situation that glaciers in Vulcan Planitia are mostly covered by water-ice debris that protects $N_2$ ice to be sublimated. (E) Syn-glaciation thrusting created north-trending arcuate ranges. The same thrusting event also affected the western margin of Vulcan Planitia. The lateral propagating north-trending thrusts were arrested by the pre-existing weak east-trending normal fault zone, which was induced by a giant impact mentioned above and marks the topographic boundary between Oz Terra and Vulcan Planitia. Impacts in Vulcan Planitia created craters surrounded by layered ejecta blankets with steep terminations. These craters reminisce rampart craters on Mars interpreted to have been generated in regions with subsurface ground-ice layers (Watters et al., 2015). (F) Deglaciation created recessive moraine landforms and deposits. Cooling of the water-ice-debris-covered $N_2$ ice created polygonal networks of fractures, while sublimation of the $N_2$ ice created pitted terrains. Finally, cooling of the debris-covered glaciers and the sublimation-induced removal of the viscous $N_2$ ice froze the sinking mega-erratic boulders derived from impact breccias formed the moated mounds.

Where the erratic boulders were sunk into the debris-covered glaciers created polygonal shaped funnel-like depressions.

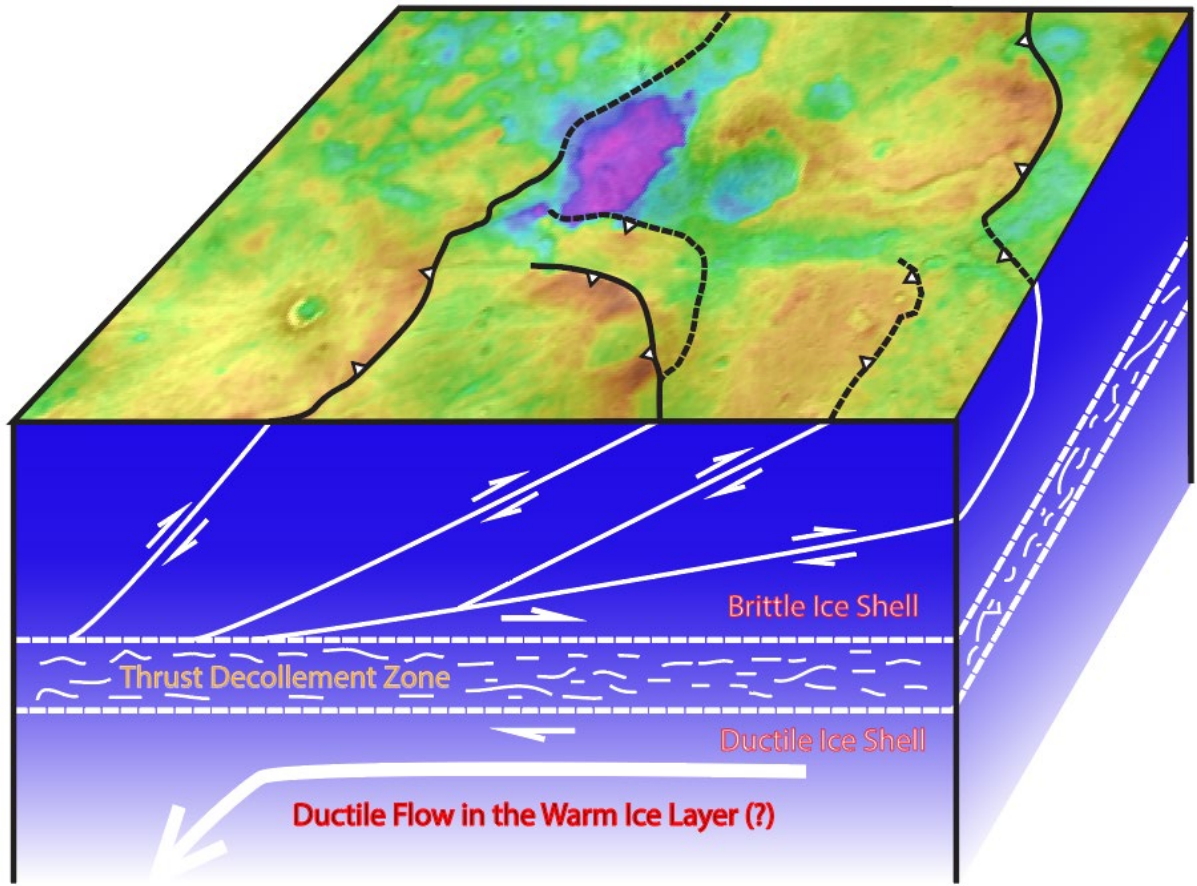

**Fig. 12. A block diagram showing the relationship between the interpreted west-dipping thrusts and a possible top-to-the-east thrust decollement zone in the brittle-ductile transition zone of Charon's ice shell.** Solid lines are thrust traces expressed as curvilinear scarps or range fronts that either truncate or offset older craters. The dashed lines are inferred thrust traces lacking fault-scarp-like morphologies. Triangles are on the hanging-wall sides of the thrusts.

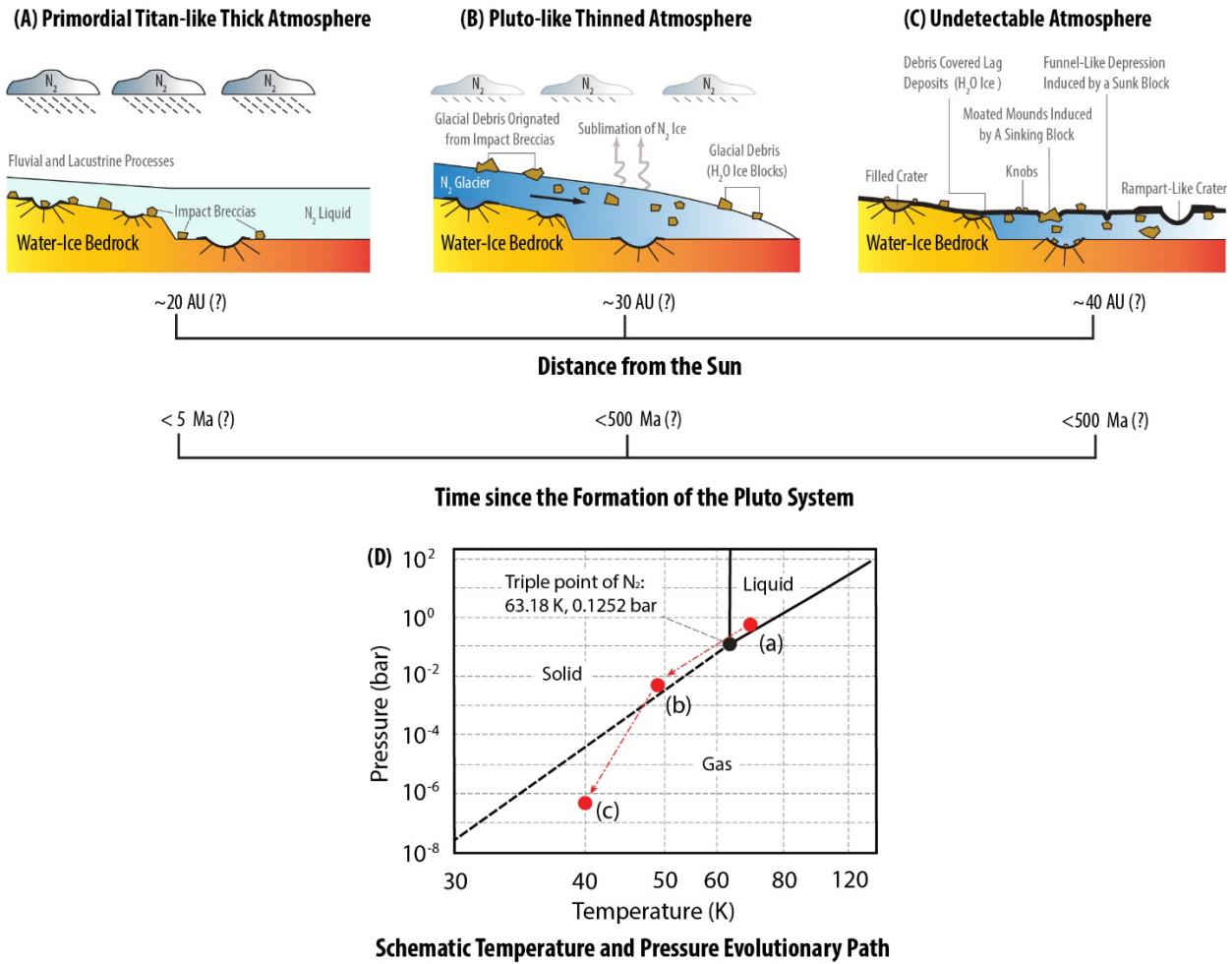

**Fig. 13. A hypothetical evolutionary path of larger Kuiper belt objects**. (A) A warm start with a thick atmosphere capable of supporting liquid $N_2$ activities. (B) Progressive thinned atmosphere cools the surface, leading to glaciation. (C) The final atmospheric loss caused rapid sublimation of $N_2$ ice. The sublimation of surface $N_2$ ice in turn causes rapid cooling, which creates a frozen landscape with a snapshot of the on-going geologic process at the time of the rapid cooling event. (D) Schematic temperature-pressure evolutionary path on the nitrogen phase diagram modified from Fray and Schmitt (2009).